\documentclass[review]{elsarticle}

\usepackage{lineno,hyperref}
\usepackage{graphicx}
\usepackage{rotating}
\usepackage{lipsum}
\usepackage{lscape}
\usepackage{amsmath}
\usepackage{lmodern}
\usepackage{natbib}
\usepackage[table,xcdraw]{xcolor}
\usepackage{textcomp}
\usepackage{float}
\usepackage{graphicx}
\usepackage[margin=1in]{geometry}
\modulolinenumbers[5]

\journal{Computer Science Review}

\begin{document}

\begin{frontmatter}

\title{Nature-Inspired Algorithms for  Wireless Sensor Networks: A Comprehensive Survey}

\author[iiserb]{Abhilash Singh}
\author[gbu]{ Sandeep Sharma \corref{san}}
\author[iitk]{Jitendra Singh}
\address[iiserb]{Indian Institute of Science Education and Research Bhopal, India}
\address[gbu]{School of ICT, Gautam Buddha University, Greater Noida, India}
\address[iitk]{Department of Electrical Engineering, Indian Institute of Technology Kanpur, India}

\cortext[san]{Corresponding author\\ Email address: sandeepsharma@gbu.ac.in (Sandeep Sharma)}


\begin{abstract}
In order to solve the critical issues in Wireless Sensor Networks (WSNs), with concern for limited sensor lifetime, nature-inspired algorithms are emerging as a suitable method. Getting optimal network coverage is one of those challenging issues that need to be examined critically before any network setup. Optimal network coverage not only minimizes the consumption of limited energy of battery-driven sensors but also reduce the sensing of redundant information. In this paper, we focus on nature-inspired optimization algorithms concerning the optimal coverage in WSNs. In the first half of the paper, we have briefly discussed the taxonomy of the optimization algorithms along with the problem domains in WSNs. In the second half of the paper, we have compared the performance of two nature-inspired algorithms for getting optimal coverage in WSNs. The first one is a combined Improved Genetic Algorithm and Binary Ant Colony Algorithm (IGA-BACA), and the second one is Lion Optimization (LO). The simulation results confirm that LO gives better network coverage, and the convergence rate of LO is faster than that of IGA-BACA. Further, we observed that the optimal coverage is achieved at a lesser number of generations in LO as compared to IGA-BACA. This review will help researchers to explore the applications in this field as well as beyond this area.
\end{abstract}

\end{frontmatter}


\section{Introduction}
Sensors in WSNs can sense, collect and transmit information together \cite{Sohrabi}. All these tasks need to be done 
effectively in order to minimize the wastage of limited sensor battery lifetime. We cannot increase the sensor lifetime by supplying external or additional energy because most of the sensors are deployed in hard-to-reach areas \cite{9261408,Akyildiz,Borges,Lu,sandeep,rahul,Yick,Muhammand,Sharma2020wireless}. Much work has been done to increase the lifetime of the sensor node. Liang et al. \cite{Liang}, have proposed, Huang algorithm, an optimal energy clustering algorithm to ensure balanced depletion of energy over the whole network which prolongs the lifetime of the system. Cardei et al. \cite{Cardei2005}, have proposed TianD algorithm, to extend the operational time of the sensors by arranging them into several maximal disjoint set covers that are activated successively. However, there exist some limitations in the above-listed algorithms. Huang algorithm is highly complex, and if data for communication is large, then it may block the channel. In contrast, the complexity of TianD algorithm is lower. However, it is unable to point out the redundant node, which is sensing redundant information.

In addition to the energy constraint, accurate sensing, and non-redundant information is a critical challenge in WSNs. In order to sense non-redundant information, the sensors need to be placed apart at a sufficient distance from each other so that the overlapping in the sensing region is minimum. However, if the sensors are placed at a more considerable distance away from each other, then it will create uncovered areas which are termed as coverage hole or blind areas. To ensure guaranteed coverage, Wang et al. \cite{Wang}, proposed a Coverage Configuration Protocol (CCP) which guaranteed coverage and connectivity with self-configuration for a wide range of applications. However, the CCP Algorithm gives underperformance if the numbers of sensors are significant.

After critically analyzing the problem of energy constraint and sensor node separation (\textit{i.e.}, node placement), we observed that there exist a trade-off between these two problems. In literature, researchers have proposed individual solutions to each problem of energy constraint and node placement but not collectively.  Keeping in view the limitations of the above-proposed solutions and instead considering the problem individually, we have combined these two problems as a multi-objective optimization problem. To balance this trade-off, we need to optimize a multi-objective optimization problem. To balance this trade-off, we need to optimize the multi-objective optimization problem. After successful optimization, we can achieve optimal coverage with less number of sensor nodes.

Several reviews are published in context to use of nature-inspired algorithms in WSNs \cite{tsai2016metaheuristics,nanda2014survey,iqbal2015wireless,demigha2012energy,kulkarni2010computational}. However, only a few cover the optimal coverage aspect in WSNs \cite{tsai2015metaheuristics,molina2008optimal,al2019evolutionary}. In \cite{tsai2015metaheuristics}, they discussed the various issues that are generally encountered while using a nature-inspired algorithm-based optimization technique for sensor deployment that leads to the optimal coverage. Whereas in \cite{molina2008optimal}, they compared three algorithms namely, standard Multi-Objective Evolutionary Algorithm (MOEA), Non-dominated Sorting Genetic Algorithm (NSGA-II)
and Indicator-Based Evolutionary Algorithms for optimal coverage in WSNs. Recently, \cite{al2019evolutionary} efficiently discussed the theoretical, mathematical and
practical application of nature-inspired algorithms in WSNs. They discussed the genetic algorithm, evolutionary deferential algorithm, NSGA and genetic programming in-depth for routing, clustering, coverage and localization in WSNs. Nevertheless, none of them provides a critical review of the problem domains in WSNs and in particular of the optimal coverage. In this paper, firstly, we have briefly discussed the nature-inspired algorithms and their application in WSNs. We have also discussed the advantages and disadvantages of the work done by various researchers. Later, we have compared the performance of two such algorithms for the optimization of a multi-objective optimization problem stated above. The first algorithm is IGA-BACA \cite{Grefenstette,sun,kk,Tian2016}. It is a hybrid of the modified evolutionary and swarm-based nature-inspired algorithm. In contrast, the second one is LO \cite{yazdani2016lion}, which is a purely swarm-based nature-inspired algorithm.

\begin{figure}[h]
\centering
\includegraphics[width=1\textwidth]{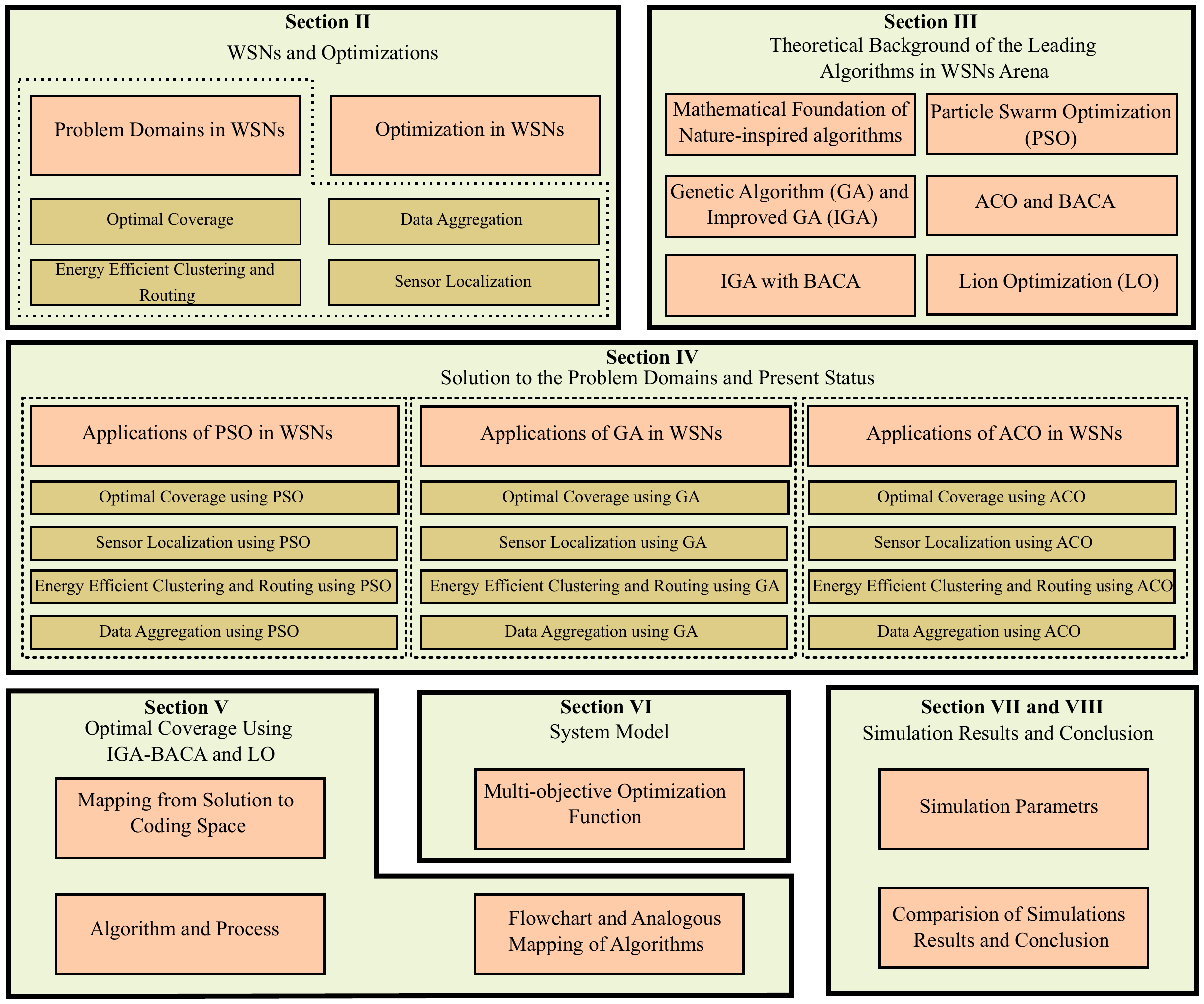}
\caption{Organization of the paper.}
\label{f:outline}
\end{figure}

The rest of the paper is organized as follows. 
In Section \ref{WSNs and Optimizations}, we have discussed the WSN’s problem domains that consist of the critical issues of the WSNs by categorizing it into four categories which are followed by a brief discussion of the taxonomy of some of the prominent optimization algorithms. Further, in Section \ref{sec:Theoretical background of the leading algorithms in WSNs arena}, we have briefly discussed the theoretical and mathematical aspect of some nature-inspired algorithms. Furthermore, in Section \ref{Solution to the problem domains and present status}, we have discussed the solutions to the problem domains of WSNs. Afterwards, in Section \ref{Optimal coverage using IGA-BACA and LO}, we have discussed the optimal coverage aspect in detail with respect to nature-inspired algorithms. Then, we have presented the system model in Section \ref{System Model}. After that, we have presented the simulation results in Section \ref{Simulation Results}. Lastly, in Section \ref{Conclusion}, we have presented the conclusion and the future scope of the work. For better readability, the outline of the paper is shown is Fig. \ref{f:outline}.

\begin{figure}[h]
\centering
\includegraphics[width=0.45\textwidth]{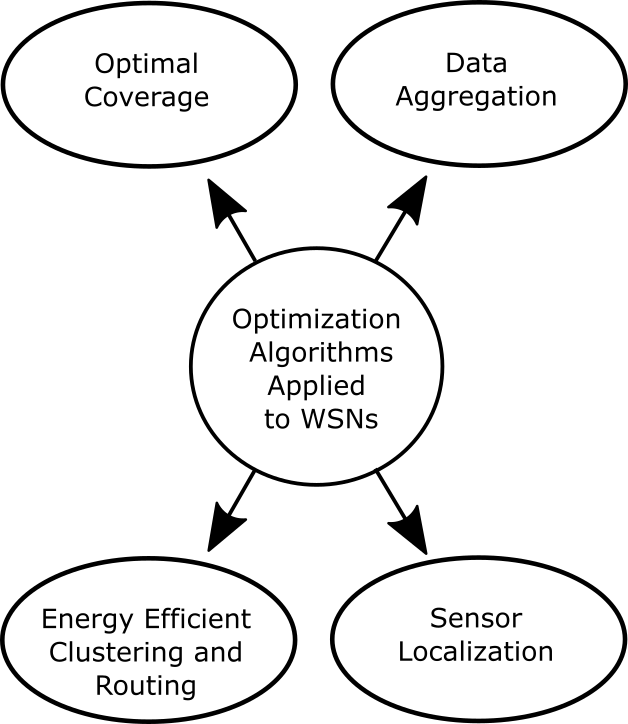}
\caption{Problem domains of WSNs.}
\label{f:application}
\end{figure}

\section{WSNs and Optimizations}
\label{WSNs and Optimizations}
The critical issues in WSNs can be broadly classified into three, namely energy efficiency, Quality of Service (QoS) and security. There exist a trade-off between all these issues. For example, if we want good QoS, then we have to compromise with the network lifetime. Same follows with the security parameters. A significant amount of work has been done concerning addressing these issues individually. However, many loopholes exist when addressing these issues individually. So, to develop a better WSNs, we need to optimize these issues simultaneously. One way of doing this is to develop a multi-objective function and optimize it by using a suitable optimizer or algorithm. The selection of suitable algorithm depends upon various factors such as the behaviour of the algorithm, type of problem, time constraint, resource availability, and desired accuracy. We have first discussed the problem domains in WSNs and then review the optimizations techniques that are available to date to solve it.

\begin{figure}[h]
\centering
\includegraphics[width=0.5\textwidth]{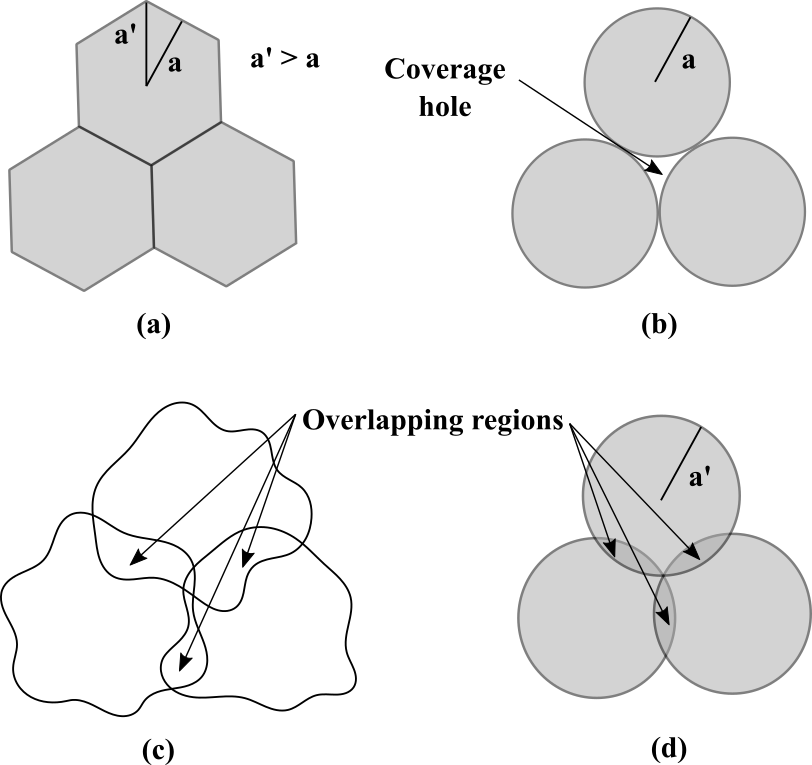}
\caption{Coverage: (a) Hexagonal shape-based,   (b) Circular 
shape-based (radius = a), (c) Real time-based 
(irregular) and (d) Circular shape-based (radius $=a^{'}$).}
\label{f:coverage}
\end{figure}
\subsection{Problem Domains in WSNs}
We have reviewed the potential of optimization and focused on the different areas in WSNs, as shown in Fig. \ref{f:application}.

\begin{itemize}
    \item Optimal Coverage in WSNs
    \item Data Aggregation in WSNs
    \item Energy Efficient Clustering and Routing in WSNs
    \item Sensor Localization in WSNs
\end{itemize}

We have first briefly discussed each of these problem domains followed by the work done to solve these issues using optimizations in the section \ref{Solution to the problem domains and present status}.   

\subsubsection{Optimal Coverage in WSNs}
Coverage is necessary and hence becomes an essential topic in the study of WSNs. The coverage in a given target area is defined as finding a set of sensors for covering the given area or all the target points. Optimal coverage means covering the entire area or all the targets point with a minimum number of sensors.

One of the crucial parameters in the coverage of a sensor in WSNs is the shape of the sensing area. In Fig. \ref{f:coverage} (a) - (d), we present four two-dimensional geometrical-based sensing shapes. In real life, the shape of the sensing area is irregular and complex due to the terrain features and solid structures. The Fig. \ref{f:coverage} (c), represents a typical example of real-life sensing shape of a sensor. However, for computational and conceptual ease, we often adopt either a hexagonal shape or a circular shape. The hexagonal shape is often applied for analysis in the WSNs because of its flexibility and no overlapping, as illustrated in Fig. \ref{f:coverage} (a). However, because of the low complexity, the circular shape is more popular. The limitation associated with the circular shape is that it creates a coverage hole, as illustrated in Fig. \ref{f:coverage} (b). This limitation is compensated by increasing the radius of the circle, as illustrated in Fig. \ref{f:coverage} (d). However, this gives birth to a new issue of overlapping regions. These overlapping regions lead to the sensing of redundant information and wastage of the limited sensor battery. However, if we critically compared all the three possibilities with the real-life sensing shape, then Fig. \ref{f:coverage} (d), comes out to be the representative of Fig. \ref{f:coverage} (c).

The only challenging issue in this problem domain is the reduction of these overlapping sensing regions with no coverage hole. The more the overlapping area, the more redundant information will be sensed by the sensors and hence more will be the wastage of the limited battery of the sensors. One way of minimizing this redundancy is to optimize the sensor node placement, which is a single-objective optimization problem. We can extend the single objective to multi-objective by considering the other network parameters.

\subsubsection{Data Aggregation in WSNs}
The second way of minimizing the sensing of redundant information is data aggregation. It is an energy-efficient technique in WSNs. While monitoring an area, sensors collect local information and send it either the complete processed data or partially processed data to the data aggregation centre. According to the data received, the data aggregation centre takes a specific decision to improve the lifetime of the sensors by eliminating the sensing of overlap or common regions.

\begin{figure*}[t!]
\centering
\includegraphics[width=0.8\textwidth]{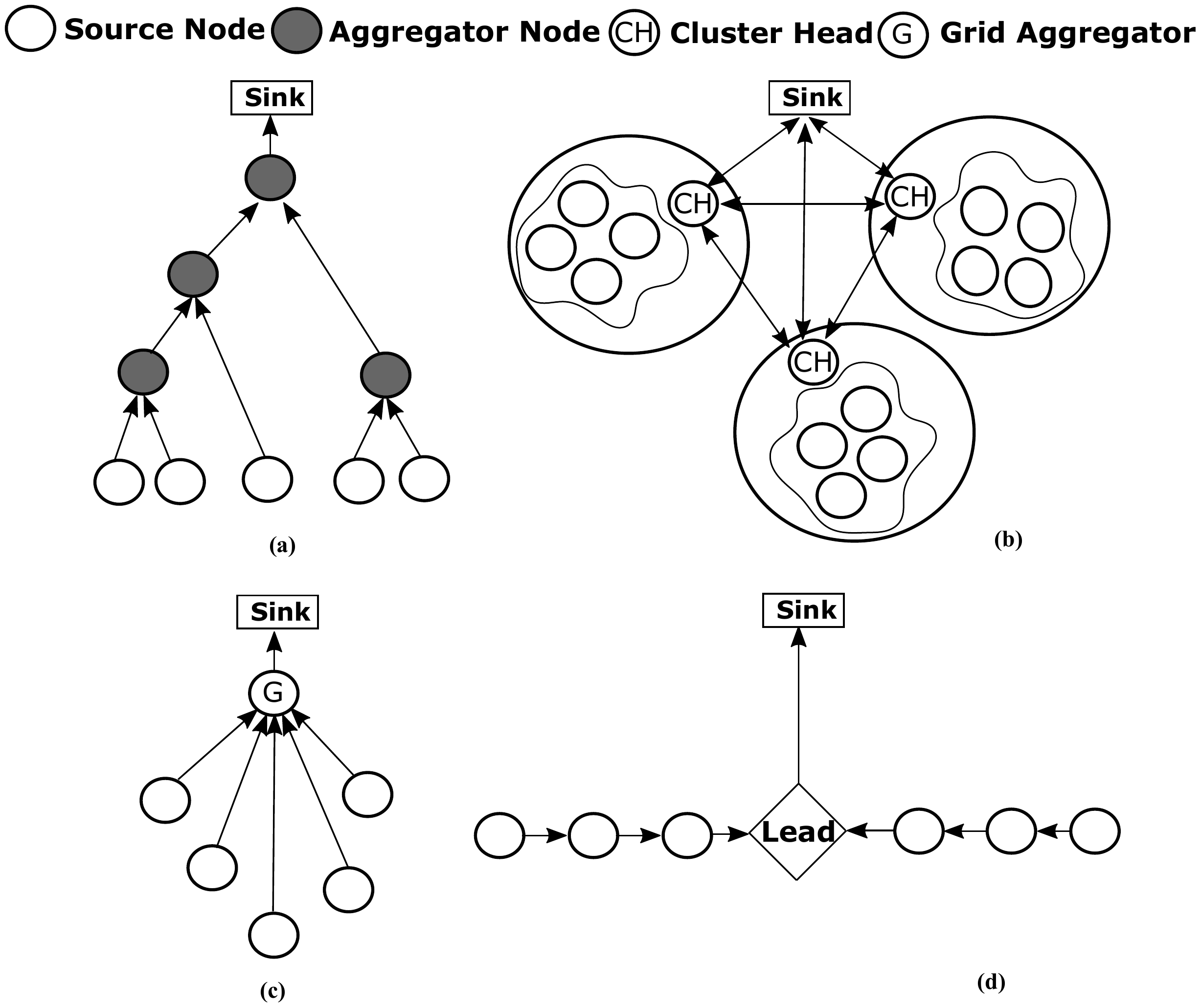}
\caption{Types of Data aggregations: (a) Tree-based,   (b) Cluster-based, (c) Grid-based and (d) Chain based.}
\label{f:dataaggre}
\end{figure*}

We can broadly classify the data aggregation techniques into four, \textit{i.e.}, Tree-based, Cluster-based, Grid-based and Chain based. All four types have been illustrated in Fig. \ref{f:dataaggre} (a) - (d). The Tree-based data aggregation technique is based on tree architecture in which the source node act as coordinator, and the data aggregation takes place at the intermediate nodes known as the aggregator node. The lower level nodes forward the information to the upper-level nodes. The Cluster-based aggregation technique is based on clustering architecture. In this type of data aggregation, the network is first divided into several clusters followed by Cluster Head (CH) selection based on sensor parameters such as sensor energy, etc. The CH first aggregates the data locally within the clusters, and then the aggregated data is sent to the sink. For each new round of data transmission, a new CH is selected to avoid excess energy consumption from the CH. In the Grid-based aggregation technique, the network is first divided into several areas, and every area reports the occurrence of any new event. The data aggregation take place at the grid aggregator node, also known as the central node. In the Chain based aggregation technique, the sensor node transfer the data to its neighbouring node and the data aggregation take place at the lead.

The main challenging issues in this problem domain are
\begin{itemize}
    \item To address the problem of optimal power allocation.
    \item Finding minimum no. of aggregation points while routing data.
    \item Perform consistency for large scale and dynamic WSNs.
\end{itemize}

\subsubsection{Energy Efficient Clustering and Routing in WSNs}
Due to the limited energy supply in sensors, the need for energy-efficient infrastructure is of utmost importance. Most of the sensor energy is consumed in the transmission of the sensed data. The energy required for the data transmission increases exponentially with the transmission length. Due to which the data transmission in sensors follows multi-hop communication. Routing in WSNs is referred to as the path traversed by the data packets to reach the sink from the source node.
First, the sensors are clustered into groups. Then a CHs is selected for each group which collects all the data from the non-CH sensors. Subsequently, the collected data is transmitted to the sink using optimal routing techniques.

The main challenging issues in this problem domain are
\begin{itemize}
    \item Selection of high energy CHs and an optimal routing path in each round.
    \item Maximization of the data delivered and the network lifetime.
    \item Communication distance minimization.
\end{itemize}

\begin{figure}[t!]
\centering
\includegraphics[width=0.5\textwidth]{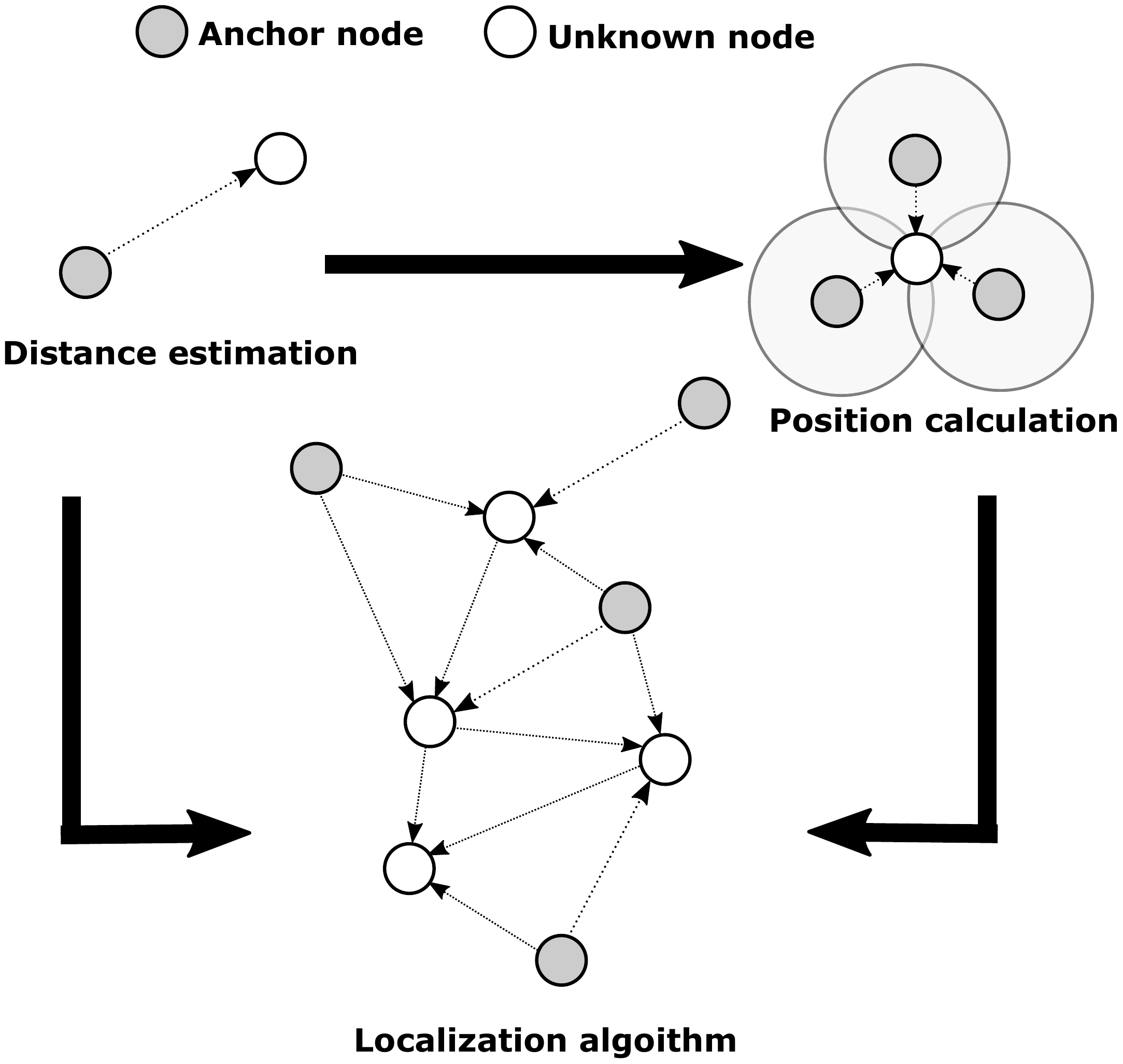}
\caption{Working of the localization system.}
\label{f:localization}
\end{figure}

\subsubsection{Sensor Localization in WSNs}
Sensor localization is the process of calculating the location of the sensor present in a network. It consists of two phases. The first one is the distance estimation, and the second one is the position calculation, as illustrated in Fig. \ref{f:localization}. The anchor or beacon node is the node with known location either through Global Positioning System (GPS) or by manual pre-programming during deployment. During the first phase, the relative distance between the anchor and the unknown node is estimated. The coordinates of the unknown node concerning the anchor nodes are calculated in the second phase using this gathered information. In order to localize the other nodes in the WSNs, the available information of distances and positions are manipulated by using various localization algorithms. A details study of such algorithms can be found in \cite{han2013localization}.

The main challenging issues in this problem domain are
\begin{itemize}
    \item Minimization of the localization error.
    \item Increasing the accuracy of the unknown node location.
\end{itemize}

\subsection{Optimization in WSNs}
\label{sec:optimization}
\begin{figure*}[h!]
\centering
\includegraphics[width=0.9\textwidth]{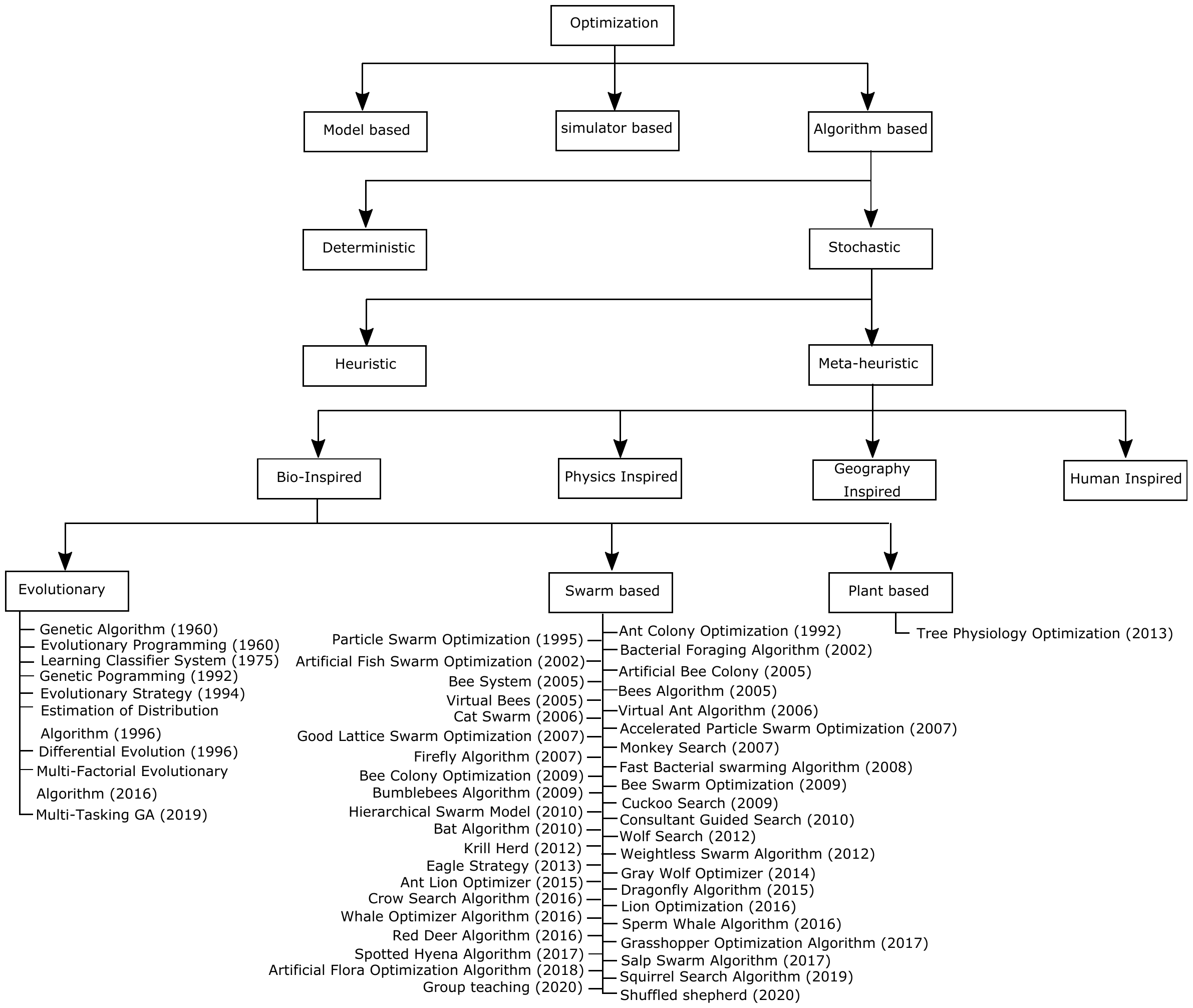}
\caption{Taxonomy of the optimization techniques.}
\label{f:taxonomy}
\end{figure*}

An optimization can be done by a model, or by a simulator or by an algorithm.  In this paper, we have evaluated the potential of optimization of the problem domains in WSNs based on algorithm approach. A detailed taxonomy of the optimization algorithms that are frequently used in WSNs is shown in Fig. \ref{f:taxonomy}. However, there exist more than 100 nature-inspired algorithms since 2000. Hence, it is not possible to list all the existing algorithms in one taxonomy. For example, Xing and Gao \cite{xing2014innovative} have listed 134 such algorithms and an online repository \textit{Bestiary} list more than 200 algorithms \cite{Bestiary}. The most recent and complete taxonomies or databases of the nature-inspired algorithms can be found in \cite{tzanetos2020comprehensive}.

The optimization algorithms are classified into deterministic (local search) and stochastic (global search). In deterministic methods, we have a theoretical guarantee of reaching the global minimum or at least to a local minimum, whereas stochastic methods only provide a guarantee in terms of probability. However, stochastic methods are faster as compared to the deterministic one. Moreover, stochastic methods are suitable for black-box formation and ill-behaved functions. In contrast to stochastic methods, the deterministic method mostly relies on the theoretical assumptions about the problem formulation and also on its analytical properties.

\begin{figure}[h!]
\centering
\includegraphics[width=0.5\textwidth]{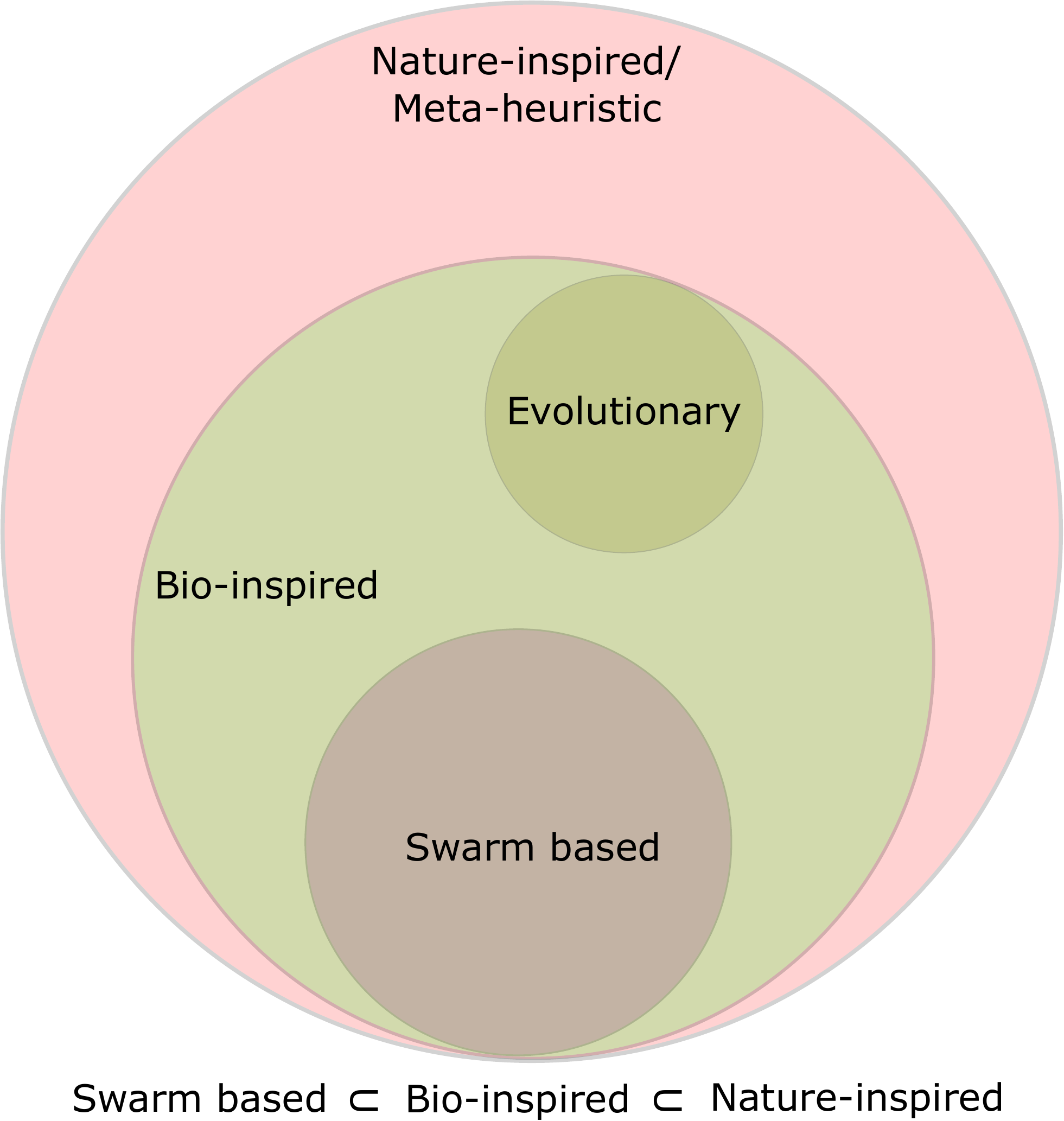}
\caption{Venn diagram for broad classification of optimization algorithm.}
\label{f:concept_swarm}
\end{figure}
\begin{figure}[h!]
\centering
\includegraphics[width=0.5\textwidth]{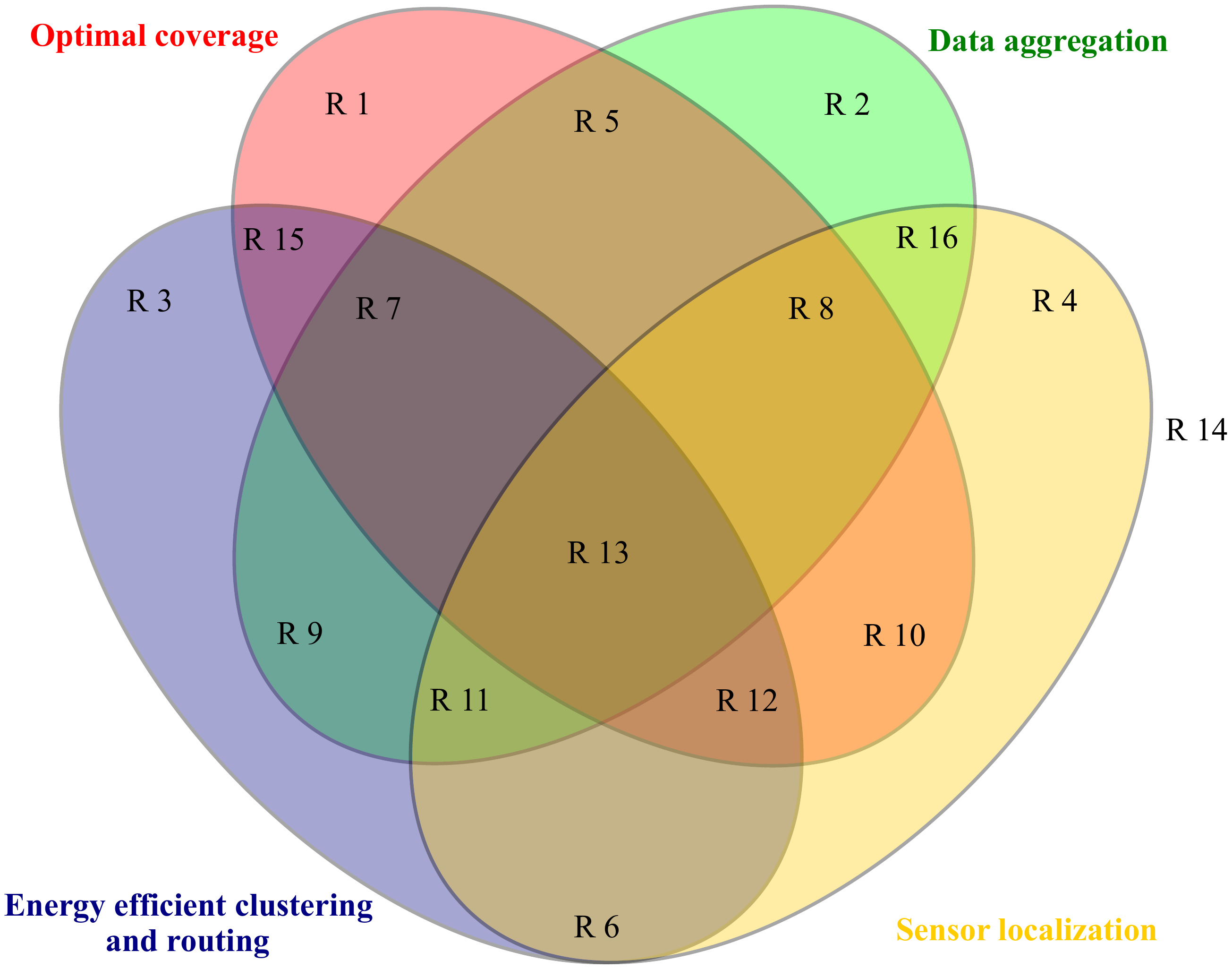}
\caption{Regions of application for various algorithm.}
\label{f:region_application}
\end{figure}

Further, the stochastic methods are classified into a heuristic and meta-heuristic algorithm. Both types of algorithms are used to increase the speed of the process of finding a global optimum for the cases where finding an optimal solution is difficult. Heuristics algorithms are problem-dependent algorithms. Due to its adaptiveness with the problem and greedy nature, they are highly prone to get stuck at local optima; resulting into failure of obtaining global optima. In contrast, meta-heuristics algorithms are problem-independent algorithms. The non-adaptive and non-greedy nature of these algorithms enables its use as a black box. These algorithms sometimes accept a temporary deterioration of the solution (\textit{e.g.},  simulated-annealing method) in order to get the global optima. The meta-heuristic algorithms are also known as nature-inspired algorithms, or intelligent optimization algorithms \cite{tao2015brief,pham2012intelligent,zhang2019general}. These algorithms are formulated by delineating inspiration from nature. The nature-inspired/ meta-heuristic algorithms are further classified as bio-inspired, physics-inspired, geography inspired and human-inspired. The majority of the nature-inspired algorithms are inspired by the biological system. Hence, a big chunk of nature-inspired algorithms are bio-inspired (biology-inspired) (Fig. \ref{f:concept_swarm}). The bio-inspired algorithms are further classified into three, namely evolutionary, swarm-based and plant-based. The evolutionary algorithms are based on the principle of evolution, such as  Darwin's principle of selection, heredity and variation \cite{dasgupta1997evolutionary}. In contrast, swarm algorithms are based on the collective intelligence \cite{kennedy2006swarm,eberhart2001swarm}.

For representing the present status of these algorithms in context to WSNs, we have created a Venn diagram (Fig. \ref{f:region_application}) that illustrate the different regions of applications or problem domains. In Fig. \ref{f:region_application}, the region R 1, R 2, R 3 and R 4 represents the application area for optimal coverage, data aggregation, energy-efficient clustering and routing and sensor localization respectively. Also, the overlapping regions have combined regions of application (\textit{e.g.}, R 3 represents the application area that includes optimal coverage as well as data aggregation). Finally, Table \ref{tab:algo_appications} represents the current status of the bio-inspired algorithms in context to WSNs.

Not all the bio-inspired algorithms are of potential use in WSNs. The algorithms for any specific problems in WSNs arena are selected based on the analogous parameters between the problem domain and the algorithm (\textit{e.g.}, Table \ref{table}) \citep{das2016metaheuristic,singhmathematical}. According to the previous studies (Table \ref{tab:algo_appications}), only three algorithms (PSO, GA, and ACO) covers all the problem domains of WSNs (\textit{i.e.} lies in region R 13). Hence, PSO, GA, ACO and their modifications such as IGA, BACA and IGA-BACA (combined meta-heuristic) are suitable for the optimizations of the problem domains in WSNs. In this study, we have evaluated the potential of the LO for optimal coverage in WSNs.

In the next section, we have tried to elaborate and give an insight into all these algorithms.

\begin{table*}[]
\centering
\caption{Algorithms with region of application. }\label{tab:algo_appications}
\resizebox*{.8\textwidth}{1.3\textwidth}{%
\begin{tabular}{ccc}
\hline
\textbf{Algorithm}                                                                                                                           & Region of application & \textbf{Main references} \\ \hline
GA \citep{holland1980adaptive}                                                                                                              & R 13                                                                         & \citep{islam2007genetic,hussain2007genetic,yoon2013efficient,peng2015improved}               \\ \hline
Evolutionary programming \citep{yao1999evolutionary}                                                                                        & R 4                                                                         & \citep{zhang2015improvement}                \\ \hline
Learning classifier system \citep{lanzi2000learning}                                                                                        & R 14                                                                         & Not addressed
\\ \hline
Genetic programming \citep{koza1997genetic}                                                                                                 & R 10                                                                         & \citep{tripathi2011wireless,aziz2016two}                \\ \hline
Evolutionary strategy \citep{back1997handbook}                                                                                              & R 10                                                                         & \citep{fayyazi2011solving,sivakumar2016performance}              \\ \hline
Estimation of distribution algorithm \citep{muhlenbein1996recombination,zhang2008rm}                                                        & R 12                                                                        & \citep{wang2012copula,cequn2011algorithm}                \\ \hline
Differential evolution \citep{qin2008differential}                                                                                          & R 12                                                                         & \citep{cui2018high,maleki2013new,kuila2014novel}              \\ \hline
Multi-factorial evolutionary algorithm \citep{gupta2015multifactorial}                                                                      & R 2                                                                         & \citep{tam2020multifactorial}                \\ \hline
Multi-tasking genetic algorithm \citep{dongrui}                                                                                             & R 14                                                                         & Not addressed                \\ \hline
ACO \citep{dorigo2010ant,dorigo2005ant,socha2008ant,blum2005ant}                                                       & R 13                                                                       & \citep{yang2010multipath,qin2010node,liao2008data,liu2014ant}              \\ \hline
PSO \citep{eberhart1995particle,shi1999empirical,kennedy1995particle}                                            & R 13                                                                        & \citep{wang2017particle,gopakumar2008localization,lu2014construction,ab2007particle}               \\ \hline
Bacterial foraging algorithm \citep{passino2002biomimicry,das2009bacterial,passino2010bacterial}                                            & R 12                                                                        & \citep{sribala2013energy,nagchoudhury2015optimal,sharma2018fuzzy}             \\ \hline
Artificial fish swarm optimization \citep{li2003studies,li2004applications}                                                                 & R 12                                                                        & \citep{song2010hierarchical,yang2016novel,yiyue2012wireless}               \\ \hline
Artificial bee colony \citep{karaboga2007artificial,karaboga2009comparative,karaboga2007powerful,karaboga2011novel} & R 12                                                                        & \citep{ozturk2012artificial,karaboga2008performance,kulkarni2016multistage,karaboga2012cluster}          \\ \hline
Bee system \citep{lucic2001bee,luvcic2003vehicle}                                                                                           & R 14                                                                        & Not addressed               \\ \hline
Bees algorithm \citep{pham2005bees,pham2006bees}                                                                                            & R 4                                                                         & \citep{moussa2010localization}                \\ \hline
Virtual bees \citep{yang2005engineering}                                                                                                    & R 14                                                                        & Not addressed                \\ \hline
Virtual ant algorithm \citep{yang2006application}                                                                                           & R 14                                                                        & Not addressed               \\ \hline
Cat swarm \citep{chu2006cat,chu2007computational}                                                                                           & R 15                                                                         & \citep{temel2013deployment,kong2014energy}                \\ \hline
Accelerated particle swarm optimization \citep{yang2011accelerated}                                                                         & R 14                                                                         & Not addressed                \\ \hline
Good lattice swarm optimization \citep{su2007good}                                                                                          & R 14                                                                        & Not addressed                \\ \hline
Monkey search \citep{mucherino2007monkey}                                                                                                   & R 3                                                                        & \citep{shankar2019integrated}                \\ \hline
Firefly algorithm \citep{yang2009firefly,yang2010eagle,yang2010firefly}                                                                   & R 12                                                                        & \citep{manshahia2016firefly,tuba2017mobile,sai2015parallel}              \\ \hline
Fast bacterial swarming algorithm \citep{chu2008fast}                                                                                       & R 14                                                                        & Not addressed                \\ \hline
Bee colony optimization \citep{davidovic2016bee}                                                                                            & R 2                                                                         & \citep{kumar2016bee}                \\ \hline
Bee swarm optimization  \citep{drias2010bees,djenouri2012bees}                                                                       & R 14                                                                         & Not addressed               \\ \hline
Bumblebees algorithm \citep{comellas2009bumblebees}                                                                                         & R 14                                                                        & Not addressed                 \\ \hline
Cuckoo search \citep{9yang2009cuckoo,3yang2013multiobjective,4yang2014cuckoo}                                                               & R 11                                                                         & \citep{goyal2014wireless,adnan2016novel,dhivya2011cuckoo}           \\ \hline
Hierarchical swarm model \citep{chen2010hierarchical}                                                                                       & R 14                                                                         & Not addressed                 \\ \hline
Consultant guided search \citep{iordache2010consultant,iordache2010consultant1,iordache2010consultant2}                                     & R 14                                                                         & Not addressed               \\ \hline
Bat algorithm \citep{yang2010new}                                                                                                           & R 12                                                                        & \citep{kaur2015radially,ng2018smart,goyal2013wireless}              \\ \hline
Wolf search \citep{tang2012wolf}                                                                                                            & R 14                                                                         & Not addressed                \\ \hline
Krill herd \citep{gandomi2012krill}                                                                                                         & R 15                                                                         & \citep{shopon2016krill,andaliby2018dynamic}                \\ \hline
Weightless swarm algorithm \citep{ting2012weightless}                                                                                       & R 14                                                                         & Not addressed                \\ \hline
Eagle strategy \citep{yang2012two}                                                                                                          & R 14                                                                        & Not addressed                 \\ \hline
Gray wolf optimizer \citep{mirjalili2015effective,mirjalili2014grey}                                                                        & R 12                                                                        & \citep{dao2016enhanced,rajakumar2017gwo,al2016grey}                    \\ \hline
Ant lion optimizer \citep{mirjalili2015ant}                                                                                                 & R 5                                                                         & \citep{yogarajan2018improved,liu2018node}                \\ \hline
Dragonfly algorithm \citep{mirjalili2016dragonfly}                                                                                          & R 6                                                                        & \citep{vinodhini2019hybrid,daely2016range}                \\ \hline
Crow search algorithm \citep{askarzadeh2016novel}                                                                                           & R 3                                                                        & \citep{mahesh2019decsa}                \\ \hline
LO \citep{yazdani2016lion}                                                                                                                  & R 9                                                                         & \citep{yogarajan2018improved,yuvaraj2019efficient}              \\ \hline
Whale optimizer algorithm \citep{mirjalili2016whale}                                                                                        & R 12                                                                         & \citep{ozdaug2017new,lang2019wireless,jadhav2017whale}            \\ \hline
Sperm whale algorithm \citep{ebrahimi2016sperm}                                                                                             & R 14                                                                        & Not addressed                \\ \hline
Red deer algorithm \citep{fard2016red}                                                                                                      & R 14                                                                        & Not addressed               \\ \hline
Grasshopper optimization algorithm \citep{mirjalili2018grasshopper}                                                                         & R 14                                                                        & Not addressed                \\ \hline
Spotted hyena algorithm \citep{dhiman2018multi}                                                                                             & R 14                                                                        & Not addressed               \\ \hline
Salp swarm algorithm \citep{mirjalili2017salp}                                                                                              & R 10                                                                        & \citep{kanoosh2019salp,syed2019weighted}                \\ \hline
Artificial flora optimization algorithm \citep{cheng2018artificial}                                                                         & R 14                                                                        & Not addressed                \\ \hline
Squirrel search algorithm \citep{jain2019novel}                                                                                             & R 14                                                                        & Not addressed                \\ \hline

Shuffled shepherd algorithm \citep{kaveh2020shuffled}                                                                                       & R 14                                                                        & Not addressed                \\ \hline
Group teaching algorithm \citep{zhang2020group}                                                                                           & R 14                                                                        & Not addressed              \\ \hline
\end{tabular}
}
\end{table*}

\section{Theoretical Background of the Leading Algorithms in WSNs Arena}
\label{sec:Theoretical background of the leading algorithms in WSNs arena}
\subsection{Mathematical Foundation of the Nature-inspired Algorithms}
In this sub-section, we have discussed the generic mathematics of nature-inspired algorithms. In computational science, any optimization algorithm can mathematically analyze in terms of an iterative process. According to \cite{yang2013swarm,yang2020nature}, any nature-inspired algorithm with $k$ parameters, $p = (p_1, ..., p_k)$, and $m$ random variables, $\epsilon = (\epsilon_1, ..., \epsilon_m)$ for a single-agent trajectory-based system can be mathematically expressed as
\begin{equation}
    x^{t=1} = \phi (x^t, p(t), \epsilon(t))
    \label{eq:mathsinglenature}
\end{equation}
where, $\phi$ represent the non-linear mapping from the current solution (at $t$) to the better solution (at $t+1$).

For population based system with $n$ swarm solution, the equation \ref{eq:mathsinglenature} can be extended to

\begin{equation}
   \begin{bmatrix}
           x_{1} \\
           x_{2} \\
           \vdots \\
           x_{n}
         \end{bmatrix}^{t+1} = \phi\Big((x^t_{1}, x^t_{2}, \cdots, x^t_{n}); (p_1, p_2, ..., p_k); (\epsilon_1, \epsilon_2, ..., \epsilon_m) \Big)
         \begin{bmatrix}
           x_{1} \\
           x_{2} \\
           \vdots \\
           x_{n}
         \end{bmatrix}^{t}
    \label{eq:mathmultinature}
\end{equation}

where, $(p_1, p_2, ..., p_k)$ represent algorithm-dependent parameters and $(\epsilon_1, \epsilon_2, ..., \epsilon_m)$ represents the random variables used for incorporating the randomization in the algorithm. This mathematical representation can include all the nature-inspired/meta-heuristic algorithm listed in Fig. \ref{f:taxonomy}.
\subsection{Particle Swarm Optimization (PSO)}
PSO was given by Kennedy and Eberhart in 1995 \citep{eberhart1995particle,kennedy1995particle}. The basic PSO was based on the simulation of the single directed, controlled motion of a swarm of flying birds. Each of these birds is treated as particles which regulate their flying information by its own and neighbour's flying experience. In other words, it combines the self-experience with the social experience, hence it was a social behaviour simulator. Later on, several revised versions of PSO emerged in which additional parameters such as confidence factors ($c_1,c_2$) and inertia weight ($w$) were added \citep{del2008particle,shi1998modified}. A recent study on PSO and its taxonomy can be found in \citep{esmin2015review}.

The initialization is random, and after that, several iterations are carried out with the particle velocity ($v$) and position ($x$) updated at the end of each iteration, as follows:
Each particle (\textit{i.e.} bird) is represented by a particle number $i$. Each particle possesses a position which is defined by coordinates in $n$-dimension space and velocity, which reflects their proximity to the optimal/best position. At first, the initialization is random, and after that, the particles are manipulated by several iterations carried out with equation \ref{eq:position} and \ref{eq:velocity} for position and velocity, respectively.
  
\begin{equation}
   x^i(k+1) = x^i(k)+v^i(k+1)
    \label{eq:position}
    \makebox[1.75cm]{}
\end{equation}
\begin{equation}
  \resizebox{0.6\hsize}{!}{$v^i(k+1) = w^iv^i(k)+c_1r_1(x^i_{best}-x^i(k))+c_2r_2(x_{gbest}-x^i(k))$}
    \label{eq:velocity}
\end{equation}
where;\\
$i$ = 1,2,...,$N_s$; $N_s$ is the size of the swarm\\
$k$ =1,2,...
$w^i$ = inertia weight for each particle i\\
$x^i_{best}$ = best location of the particle\\
$x_{gbest}$ = best location amongst all the particle in swarm\\
$c_1$ = confidence factor which represents the private thinking of the particle itself; assigned to $x^i_{best}$\\
$c_2$ = confidence factor which represents the collaboration among the particles; assigned to $x_{gbest}$\\
$r_1, r_2$ = random values between [0 ,1].

\subsection{GA and Adaptive GA(or IGA)}
John H. Holland and his collaborators proposed the genetic algorithm in the 1960s and 1970s \citep{holland1980adaptive}, and since then it has become one of the widely used meta-heuristic algorithms. It is based on the abstraction of Darwin’s evolution principle of biological systems that has three components or genetic operators; reproduction-crossover-mutation. Every solution is encoded in a string (often decimal or binary) called chromosomes. The fitness function in every iteration calculates its value. Afterwards, these values are sorted in descending order. Solutions that are present at the top are considered as good solutions and selected for reproduction. It discards the solutions with low fitness values. After completion of reproduction, the selected solutions will go through crossover and mutation. The role of the crossover operator is to produce crossed solutions with optimal fitness values by the interchange of genetic material. The probability of this event is known as crossover probability, represented by ${P_c}$. This event is followed by mutation; which targets to find the unexplored genetic material with a probability known as mutation probability, represented by ${P_m}$. The computational equations for ${P_c}$ and ${P_m}$ is given by \ref{eq:1} and \ref{eq:2}.

\begin{equation}\label{eq:1}
P_c =\begin{cases}
\frac{k_1(f_{max}-f')}{f_{max}-f_{avg}} & f'>f_{avg} \\ k_3 & f'<f_{avg}
\end{cases} 
\makebox[1cm]{}
\end{equation}
\begin{equation}\label{eq:2}
P_m =\begin{cases} \frac{k_2(f_{max}-f)}{f_{max}-f_{avg}} & f>f_{avg} \\
k_4 & f<f_{avg}
\end{cases} 
\makebox[1.75cm]{}
\end{equation}

\begin{figure}
\centering
\includegraphics[width=0.5\textwidth]{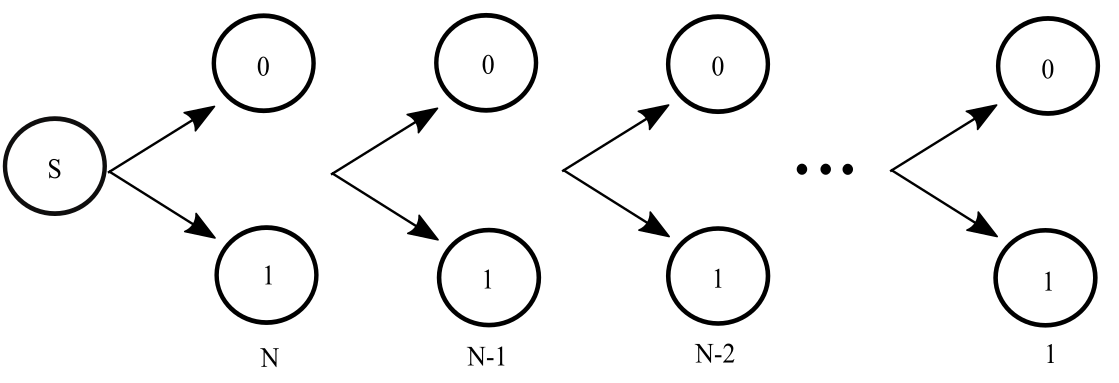}
\caption{BACA network.}
\label{f:1}
\end{figure}
Where ${f_{max}}$ and ${f_{avg}}$  represents the highest fitness and average fitness of the population respectively, ${f'}$ represents the higher fitness amongst the two solutions that are selected for crossover and ${f}$ represents the fitness of the solution that is selected for mutation. In order to restrict  the values of ${P_c}$ and ${P_m}$ in the range [0,1], the values of the constants ${k_1, k_2, k_3 }$ and ${k_4}$ should be less than ${1}$. Also, the constants $k_1$ and $k_3$ should be greater than the $k_2$ and $k_4$.

Adaptive GA (AGA) or Improved GA (IGA) is an enhanced version of conventional GA. In AGA, ${P_c}$ and ${P_m}$ changes adaptively based on different individuals condition that ultimately retard the possibility of premature convergence \citep{pal2017genetic}.

\subsection{ACO and BACA}
ACO is based on the food searching process by ants. In this process, the ant emits pheromone in the path. The remaining ants follow the path with a high intensity of pheromones \citep{wangfault,sun2010wsn}. ACO estimates the optimal path through continuous accumulation and pheromones release process in several iterations. The performance of ACO depends strongly on the early stage pheromones. Lack of sufficient pheromones may result in premature convergence (\textit{i.e.}, local optima) \citep{xiong2006binary} and to avoid this, we use BACA. A typical example of how BACA works is illustrated in Fig. \ref{f:1}. This binary coding increases the efficiency of the algorithm \citep{xiong2006binary}. Different ants search the same routine and emit the pheromones on each edge. Out of the two binary edges, each ant selects one. This process can be represented in the form of a matrix with only (2 $\times$ $n$)'s space. Defining a digraph ${G=(V,  R)}$ with $V$ representing the node-set and $R$ representing the path set.


\begin{equation}\label{eq:3}
\begin{aligned}
   \big\{ \big\{v_0(c_s),v_1((c^0_N)),v_2((c^1_N)),v_3((c^0_{N-1})),v_4((c^1_{N-1})),...,
 \\ v_{2N-3}((c^0_{2}),v_{2N-2}((c^1_{2}),v_{2N-1}((c^0_{1}),v_{2N}((c^1_{1})\big\} \big\}
 \end{aligned}
\makebox[1.75cm]{}
\end{equation}

In this digraph, ${c_s}$ represent the staring node while ${c^0_j}$  and ${c^1_j}$ represents the value 0 and 1 of ${b_j}$ bit used in the binary mapping. $N$ is the encoding length (binary). For each node present in the node set, ${(j=1, 2, 3,..., N)}$, there exist only two paths (0 and 1 states) which points towards ${c^0_{j-1}}$  and ${c^1_{j-1}}$ respectively \citep{Tian2016}. Initially, it has been assumed that all the path have same piece information
(equal to ${\tau_{i,j}(0)=C}$; $C$ is constant and ${\Delta\tau_{i,j}(0)(i,j=1, 2, ..., N)}$). During the path deciding phase, the pheromones realized by $k$ ($k$ = 1, 2, 3, ..., $m$; $m$ is the number of ants) ants and the probability of movement decides the direction. The probability of movement, ${p^k_{i,j}}$, is defined as 

\begin{equation}\label{eq:4}
\begin{aligned}
p^k_{i,j}= \frac{( \tau ^ \alpha _{i,j}(t). \eta ^ \beta _{i,j}(t))}{ (\sum_{s\in allowed k}  \tau ^ \alpha _{i,j}(t). \eta ^ \beta _{i,j}(t)) }  
\end{aligned}
\makebox[1.75cm]{}
\end{equation}

$k$ ants moves from point $i$ to point $j$. ${\alpha}$ and ${\beta}$ are constants.  ${\tau_{i,j}}$ and ${\eta_{i,j}}$ represents the unutilized information and visualness respectively in the ($i$, $j$) junction at $t$ moment.
$ {Allowed _k={(0,1)}}$ represents the upcoming status. With time, the pheromones evaporate resulting in loss of information. $\rho$ is the perseverance factor and (1-$\rho$) represents the information loss factor. The ${\tau_{i,j}}$ for the next moment is represented by 
\begin{equation}\label{eq:5}
\begin{aligned}
 \tau_{i,j}(t+1)= \rho . \tau_{i,j}(t)+ \Delta \tau_{i,j}  
\end{aligned}
\makebox[1.75cm]{}
\end{equation}

\begin{equation}\label{eq:5x}
\begin{aligned}
 \Delta \tau_{i,j}  = \frac{1}{f_{best}(S)}
\end{aligned}
\makebox[1.75cm]{}
\end{equation}
Where,

 ${f_{best}(S)}$ is the optimal cost. In a nutshell, BACA differs with the conventional ant colony only in the way the ant selects the path.

\subsection{IGA with BACA}
The combined meta-heuristic, IGA-BACA, searches for the optimal solution by initializing the BACA network with the final result of the IGA. Firstly the IGA is used to optimized the randomly generated solution. Now for the same time been, this optimized solution is feed to initialize the pheromones information of the BACA algorithm. The IGA-BACA algorithm terminates the loop once it meets the termination condition; otherwise, the complete process repeats itself to meet the termination condition (\textit{i.e.}, optimal updated pheromones).

\subsection{LO}
There are two types of lions; residents and nomads. Resident lions always live in groups called pride. In general, a pride of lion typically involves about five female along with their cubs of both sexes and one or more than one adult male. Young males, when getting sexually mature, get excluded from their birth pride. Nomads move either in pair or singularly. Pairing occurs among related males who have been excluded from their maternal pride. The lion may switch lifestyles means nomad at any time become a resident and vice versa \citep{yazdani2016lion}.

Unlike that of cats, lions hunt together to catch their prey, which increases the probability of success of hunting. In case if a prey manages to escape then the new position of prey, ${PREY'}$ is given by

\begin{equation}\label{eq:6}
\begin{aligned}
\noindent
\scalebox{1}{
\noindent
\scalebox{0.9}{
$PREY'=PREY+rand(0,1).PI.(PREY-Hunter)$ }}
\end{aligned}
\makebox[1.75cm]{}
\end{equation}

 Where  ${PREY}$ represents the current position of prey, $PI$ is the percentage of improvement in the fitness of hunter. The formulas are proposed to mimic encircling prey by the hunter group. The new positions, according to the location of prey, are generated as follows:
\begin{equation}\label{eq:7}
\begin{aligned}
\noindent
\scalebox{0.6}{
$Hunter'=\begin{cases}rand((2*PREY-Hunter),PREY) & (2*PREY-Hunter)<PREY\\rand(PREY,(2*PREY-Hunter)) & (2*PREY-Hunter)>PREY\end{cases}$ }
\end{aligned}
\makebox[1.75cm]{}
\end{equation}
 Where Hunter is the current position of the hunter. And the new position for centre hunters is given as
  \begin{equation}\label{eq:8}
  \begin{aligned}
  \noindent
  \scalebox{0.9}{
  $ Hunter'=\begin{cases}rand(Hunter,PREY) & Hunter<PREY\\rand(PREY,Hunter) & Hunter>PREY\end{cases}  $ }
  \end{aligned}
  \makebox[1.75cm]{}
  \end{equation}
  In the equation \ref{eq:7} and \ref{eq:8} ${rand(a,b)}$, generates a random number between ${a}$ and ${b}$, where ${a}$ and ${b}$ are upper and lower bound respectively. A detail of the process involve in LO is mention in the pseudo code of the literature \citep{yazdani2016lion}.

\begin{figure*}
\centering
\includegraphics[width=0.9\textwidth]{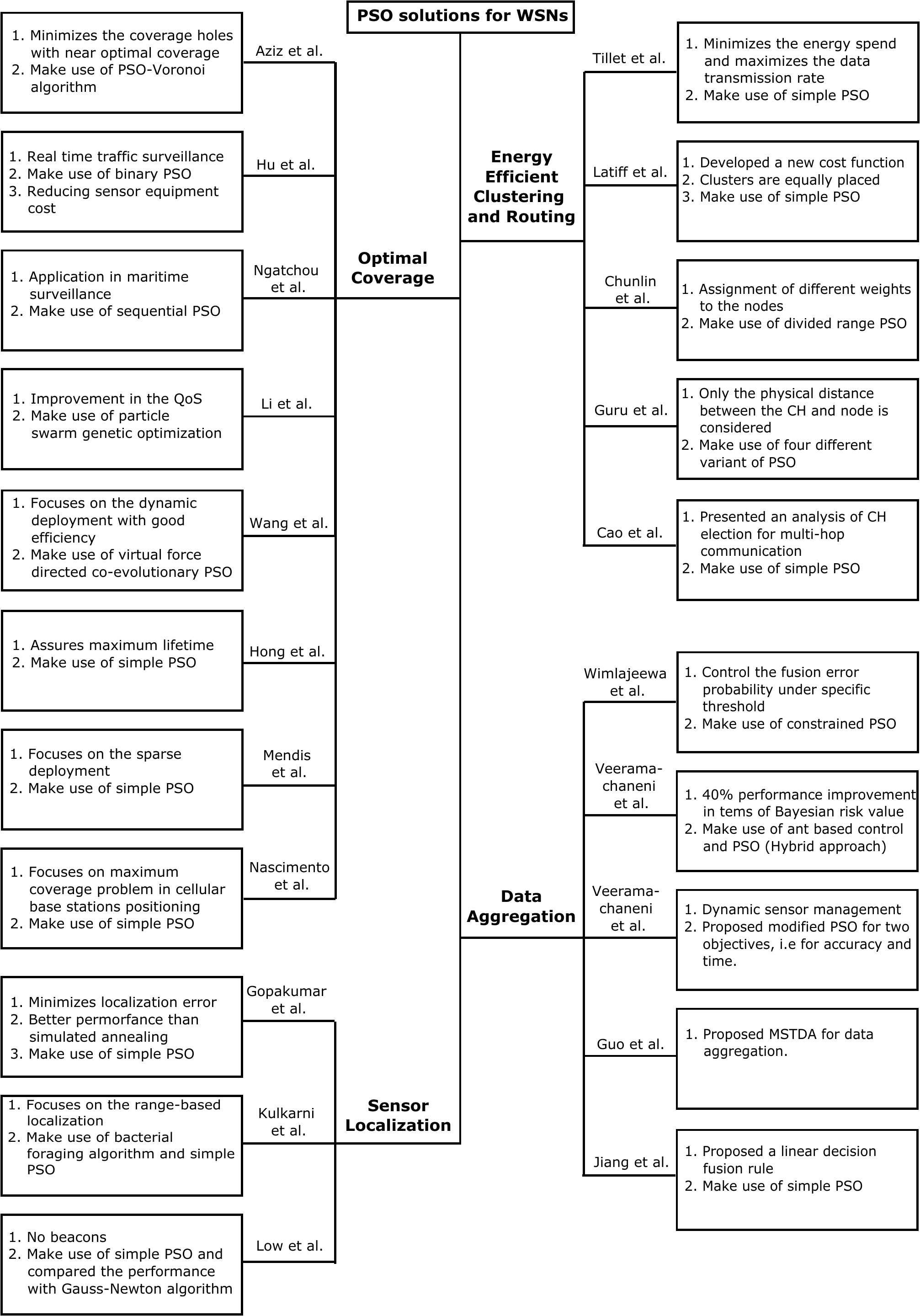}
\caption{Summary of the PSO approaches/solutions to the problem domains in WSNs.}
\label{f:psosolution}
\end{figure*}

\section{Solution to the Problem Domains and Present Status}
  \label{Solution to the problem domains and present status}

 In this section, we have presented a summary of the most prominent solutions to the problem domains of the WSNs based on some of the bio-inspired meta-heuristic algorithms, namely PSO, GA, and ACO.

 \subsection{Applications of PSO in WSNs}
  The centralized nature of PSO enables its application in the minimization of the coverage holes for near-optimal coverage in WSNs \citep{ab2007particle,ab2009wireless,hu2008topology,ngatchou2005distributed,li2007improving,wang2007improved,hong2007allocating,mendis2006optimized,nascimento2010particle}. Data aggregation is a repetitive process, which makes it suitable for PSO \citep{wimalajeewa2008optimal,veeramachaneni2008swarm,veeramachaneni2004dynamic,guo2011multi,jiang2012linear}. PSO is suitable for selecting CH's with high energy in each round \citep{guru2005particle,cao2008cluster,tillett2003particle,latiff2011extending,ji2004particle}. It also minimizes the sensor node localization errors \citep{gopakumar2008localization,kulkarni2009bio,low2008particle}. A detailed chart, illustrating the PSO based solution, is presented in Fig. \ref{f:psosolution}. 
 \subsubsection{For Optimal Coverage using PSO}

 Various studies have been reported to improve sensor coverage using PSO.
 \citet{mendis2006optimized} used the conventional PSO for optimization of the mobile sink node location in WSNs. To deal with the various complexities and challenges in different applications, various modified or improved version of PSO are proposed in the literatures.   \citet{ngatchou2005distributed} used a modified version of PSO, namely sequential PSO for distributed sonar sensor placement. Sequential PSO is generally used for high dimension optimization and found application in underwater sensor deployment optimization.  Further, \citet{li2007improving} also used a modified version of PSO, namely Particle Swarm Genetic Optimization (PSGO) for optimal sensor deployment. PSGO involve selection and mutation operator of GA, which eliminates the premature convergence issue of PSO.
 Afterwards, \citet{wang2007improved} proposed a virtual force directed co-evolutionary PSO (VFC-PSO) for dynamic sensor deployment in WSNs.
 \citet{ab2007particle} has proposed a novel optimization approach combining PSO and Voronoi diagram for sensor coverage problem in WSNs. This algorithm works efficiently for small Region of Interest (ROI) with a high number of sensor node or vice-versa. Subsequently, \citet{hong2007allocating} used conventional PSO for searching the near-optimal Base Station (BS)  location in WSNs. \citet{hu2008topology} proposed a methodology for optimal deployment of large radius sensors. They have used PSO for optimization of the sensor deployment in order to reduce the links in the proposed topology. \citet{nascimento2010particle} used the conventional PSO for BS positioning to avoid the overlapping between cells. Overall, various modified versions, including the conventional PSO, can be used for improving the sensor coverage. 
\subsubsection{For Sensor Localization using PSO}
For accurate node localization, \citet{low2008particle} used the conventional PSO. They have reported better accuracy when the results are compared with the Gauss-Newton algorithm. Similarly, \citet{gopakumar2008localization} used the same conventional PSO and reported better accuracy as compared with the simulated annealing approach. Later, \citet{kulkarni2009bio} presented a comprehensive study on node localization. They have compared the results of PSO and bacterial foraging algorithm. They reported that the node localization in WSNs is faster with PSO and more accurate with bacterial foraging algorithm.  
\subsubsection{For Energy Efficient Clustering and Routing using PSO}
Several studies have reported the use of PSO for energy-efficient clustering and routing. \citet{tillett2003particle} have used the conventional PSO for sensor node clustering in WSNs. They reported that the PSO outperforms simulated annealing and random search algorithm in terms of energy-efficient clustering. Afterwards, \citep{ji2004particle} proposed a divided range PSO algorithm for network clustering. They reported that the proposed algorithm is efficient when then mobile sensors are dense. Subsequently, \citet{guru2005particle} proposed four variants of PSO and applied it for energy-efficient clustering. They reported that the PSO with the supervisor-student model outperforms the other three algorithms. \citet{cao2008cluster} have used a hybrid of graph theory and PSO algorithm for energy-efficient clustering in multi-hop WSNs. \citet{latiff2011extending} have used the conventional PSO for re-positioning of BS in a clustered WSNs. Overall, the use of PSO reduces energy consumption and extend the network lifetime.

\subsubsection{For Data Aggregation using PSO}
\citet{veeramachaneni2004dynamic} have used the conventional PSO for optimization of the accuracy and time from data aggregation aspect. In general, they have evaluated the potential of the PSO for multi-objective optimization. Further, \citet{wimalajeewa2008optimal} have used the constrained PSO for optimal power allocation. Afterwards, \citet{veeramachaneni2008swarm} have used the hybrid version of PSO, namely ACO and PSO for dynamic sensor management. \citet{guo2011multi} proposed a multi-source temporal data aggregation algorithm (MSTDA) for data aggregation in WSNs. Subsequently, \citet{jiang2012linear} have used the constrained PSO with the penalty function concept, which increases the accuracy. 

\subsection{Applications of GA in WSNs}
\begin{figure*}
\centering
\includegraphics[scale=0.7]{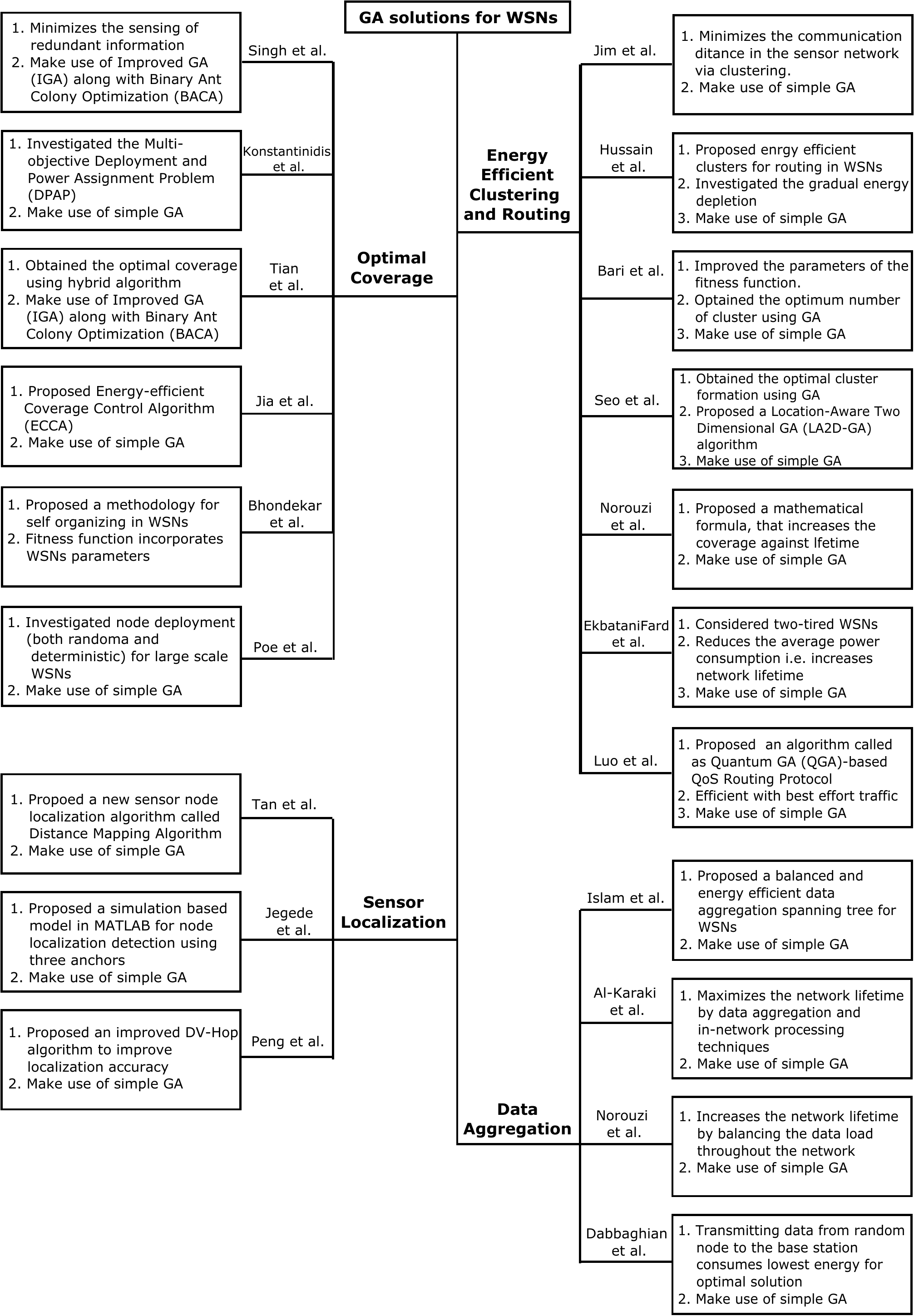}
\caption{Summary of the GA approaches/solutions to the problem domains in WSNs.}
\label{f:gasolution}
\end{figure*}

GA is proven to be good for random as well as deterministic deployment \citep{singhmathematical,Tian2016,jia2009energy,konstantinidis2008evolutionary,bhondekar2009genetic,poe2009node}. It is also good at finding lesser number of data aggregation points while routing the data to the base station \citep{islam2007genetic,al2009data,norouzi2012tree,Dabbhaghian}. It is used for pre-clustering which reduces the resultant communication distance \citep{jin2003sensor,hussain2007genetic,hussain2009genetic,seo2009evolutionary,norouzi2011new,bari2009genetic,ekbatanifard2010multi,luo2010quantum}. Besides this, the global searching capability of the GA results into higher accuracy in locating the sensor nodes \citep{peng2015improved,jegede2013genetic,tan2019distance}. A detail chart, illustrating the GA based solution is presented in Fig. \ref{f:gasolution}.

 \subsubsection{For Optimal Coverage using GA}
 Various studies have evaluated the potential of GA for network coverage optimization. \citet{konstantinidis2008evolutionary} have modeled the sensor deployment and power assignment as a multi-objective problem and used the conventional GA for the optimization.  \citet{poe2009node} proposed an approach for sensor deployment over a large WSNs. They make use of conventional GA. They have compared and reported the pros and cons of three different types of deployment. \citet{bhondekar2009genetic} have used the conventional GA for node deployment in a fixed WSNs. \citet{jia2009energy} proposed an energy-efficient novel network coverage approach using conventional GA.
 They reported that the proposed approach results in balanced performance with high network coverage rate. \citet{Tian2016} have used a hybrid version of GA called Improved GA and Binary ACO Algorithm (IGA-BACA) for optimal coverage in WSNs and compared there results with conventional GA. They reported that IGA-BACA outperforms conventional GA. They also reported a high coverage rate. Recently, \citet{singhmathematical} have used the same IGA-BACA and conventional GA for optimal coverage in WSNs with reduced sensing of redundant information.
 
 \subsubsection{For Sensor Localization using GA}
 \citet{jegede2013genetic} have used the conventional GA for node localization in WSNs. Recently, \citet{peng2015improved} have used DV-Hop GA based algorithm for node localization WSNs. They reported that the DV-Hop GA based algorithm outperforms the previously proposed algorithm. More recently, \citet{tan2019distance} have proposed Distance Mapping Algorithm (DMA) and integrate this with the GA for accurate node localization in WSNs. They reported that the proposed algorithm outperforms previously proposed algorithms in terms of accuracy and energy consumption.
 
\subsubsection{For Energy Efficient Clustering and Routing using GA}
\citet{jin2003sensor} proposed a sensor network optimization framework for 
\citet{bari2009genetic} have used the conventional GA algorithm for energy-efficient clustering and routing in a two-tier sensor network. They have reported that the proposed approach shows significant improvement compared to the earlier proposed schemes. \citet{seo2009evolutionary} proposed a hybrid GA algorithm, namely Location-Aware 2-D GA (LA2D-GA) for efficient clustering in WSNs. They reported that the LA2D outperform its 1-D version. \citet{hussain2009genetic} proposed an energy-efficient clustering and routing scheme based on conventional GA. Further, \citet{luo2010quantum} proposed the first quantum GA based QoS routing protocol for WSNs. Also, \citet{ekbatanifard2010multi} proposed a multi-objective GA for energy-efficient QoS routing approach in WSNs. They reported that the proposed approach successfully reduces the average power consumption by efficient optimization of the network parameters. \citet{norouzi2011new} proposed a dynamic clustering algorithm for WSNs based on conventional GA.

\subsubsection{For Data Aggregation using GA}
 \citet{islam2007genetic} have proposed an energy-efficient balanced data aggregation tree algorithm based on GA. They reported that the spanning tree-based proposed algorithm improves the network lifetime significantly. \citet{al2009data} have proposed a grid-based data aggregation and routing scheme for WSNs. They reported that the proposed scheme reduces power consumption and improves network lifetime. Similar to the \citet{islam2007genetic}, \citet{Dabbhaghian} proposed an energy-efficient balanced data aggregation using spanning tree and GA. They also, reported an increase in the network lifetime. However, an improved version of spanning tree-based data aggregation algorithm is proposed by \citet{norouzi2012tree}. They use the residual energy of the nodes to further improve the network lifetime.
 
 \begin{figure*}
\centering
\includegraphics[scale=0.7]{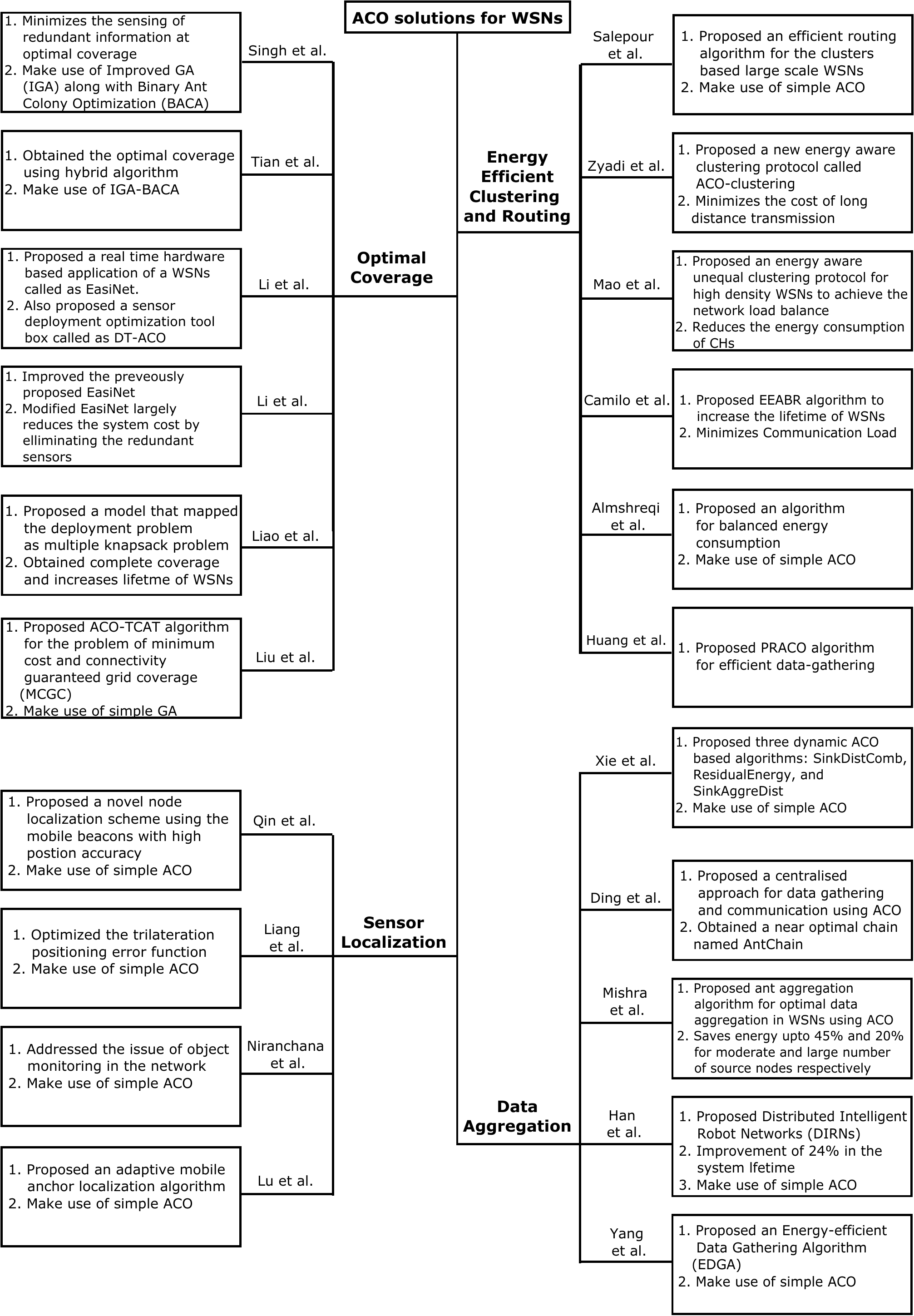}
\caption{Summary of the ACO approaches/solutions to the problem domains in WSNs.}
\label{f:acosolution}
\end{figure*}

\subsection{Applications of ACO in WSNs}
The distributed nature of ACO results in better dynamic deployment of the sensor node for near-optimal coverage \citep{singhmathematical,Tian2016,li2008demonstration,li2010easidesign,liao2011ant,liu2012sensor}. ACO performs better in case of large and dynamic WSNs \citep{ding2004data,misra2006ant,han2008maximum,inproceedingsyang,xie2012ant}. It also increases the network lifetime \citep{camilo2006energy,almshreqi2012improved,huang2010energy,salehpour2008energy,ziyadi2009adaptive,mao2013improved}. Never the less it also increases the accuracy of the unknown node in WSNs \citep{qin2010node,liang2010wireless,niranchana2012object,lu2014adaptive}. A detail chart, illustrating the ACO based solution is presented in Fig. \ref{f:acosolution}.

 \subsubsection{For Optimal Coverage using ACO}
 \citet{li2008demonstration} have proposed an efficient sensor deployment optimization toolbox named as DT-ACO. Also, they have proposed a real-time hardware-based application for WSNs called EasiNet. Later, in \citet{li2010easidesign}, they have modified the previously proposed EasiNet. This modification allows them to eliminate redundant sensors during sensor deployment. \citet{liao2011ant} have proposed an efficient approach for sensor deployment using ACO. They have formulated the deployment problem as multiple knapsack problem (MKP). They reported a complete network coverage with prolong network lifetime. \citet{liu2012sensor} proposed a novel approach for sensor deployment in WSNs using ACO with three ant transition concept. They report a high coverage rate.
 Recently, \citet{Tian2016} have used the hybrid version of ACO, namely IGA-BACA. They reported a high network coverage rate with high network lifetime. More recently, \citet{singhmathematical} have used the IGA-BACA for reducing the sensing of the redundant information with optimal coverage.

 \subsubsection{For Sensor Localization using ACO}
 \citet{qin2010node} have proposed a novel node localization scheme using ACO through beacons signals. They reported a high localization accuracy with low power consumption. \citet{niranchana2012object} have proposed a node localization approach in which the prediction of the nodes is made through interval theory and relocation of the nodes are done through ACO. They also reported a high localization accuracy. Further, \citet{liang2010wireless} have used the simple ACO for node localization in WSNs. They have optimized the trilateration positioning error function. They reported a higher-order localization accuracy compared to the previously proposed localization schemes. Recently, \citet{lu2014adaptive} have proposed an ACO based mobile anchor node localization scheme for WSNs.
 
\subsubsection{For Energy Efficient Clustering and Routing using ACO}
\citet{camilo2006energy} have proposed a new routing algorithm for WSNs based on ACO. They reported a low communication load with low energy consumption. Further, \citet{salehpour2008energy} also proposed a new routing algorithm with two routing levels based on ACO. They reported a relatively low power consumption and more load balancing.  \citet{ziyadi2009adaptive} proposed an energy-aware clustering protocol based on ACO clustering for WSNs. They reported an increase in the network lifetime. Later, \citet{huang2010energy} proposed a Prediction routing algorithm based on ACO. It was first of its kind. They reported various advantages such as low power consumption, increase in network lifetime, and high load balancing. \citet{almshreqi2012improved} proposed a self-optimization algorithm based on ACO for balance energy consumption in WSNs. They reported low energy consumption with reduced packet loss.
\citet{mao2013improved} proposed a fuzzy-based unequal clustering algorithm. They have used ACO for energy-aware routing. They reported that the proposed algorithm outperforms various traditional algorithm such as LEACH. 

\subsubsection{For Data Aggregation using ACO}
\citet{ding2004data} proposed an efficient self-adaptive data aggregation algorithm for WSNs based on ACO. They reported that the proposed algorithm outperforms the benchmark algorithms such as LEACH and PEGASIS in terms of prolonging the network lifetime. Further, \citet{misra2006ant} proposed an approach for efficient data aggregation algorithm for WSNs. They reported that the proposed approach is energy-efficient. \citet{han2008maximum} proposed a novel approach for multi-media data aggregation in wireless sensor and actor-network. They have compared the performance with the traditional methods such as MEGA and reported an improvement in the stability, accuracy and network lifetime. \citet{inproceedingsyang} proposed an energy-efficient data aggregation algorithm based on ACO for WSNs. They reported an improvement over network lifetime. Similarly, \citet{xie2012ant} proposed a data aggregation approach for WSNs using ACO and reported an improvement over network lifetime.

\begin{table*}[]
\centering
\caption{Summary of the present status of the bio-inspired algorithms approach to the problem domains of WSNs. }\label{tab:present status}
\begin{tabular}{cccccc}
\hline
\cellcolor[HTML]{9B9B9B}\textbf{\begin{tabular}[c]{@{}c@{}}Problem\\ Domians of WSNs\end{tabular}} & \textbf{\begin{tabular}[c]{@{}c@{}}Optimization\\ Algorithms\end{tabular}} & \textbf{PSO}                     & \textbf{GA}                      & \textbf{ACO}                     & \textbf{LO}                                                                                \\ \hline
\multicolumn{2}{c}{\cellcolor[HTML]{9B9B9B}\textbf{\begin{tabular}[c]{@{}c@{}}Optimal \\ Coverage\end{tabular}}}                                                                & {\color[HTML]{00009B} Addressed} & {\color[HTML]{00009B} Addressed} & {\color[HTML]{00009B} Addressed} & {\color[HTML]{3166FF} \begin{tabular}[c]{@{}c@{}}Addressed\\ (in this paper)\end{tabular}} \\ \hline
\multicolumn{2}{c}{\cellcolor[HTML]{9B9B9B}\textbf{\begin{tabular}[c]{@{}c@{}}Data \\ Aggregation\end{tabular}}}                                                                & {\color[HTML]{00009B} Addressed} & {\color[HTML]{00009B} Addressed} & {\color[HTML]{00009B} Addressed} & {\color[HTML]{00009B} Addressed}                                                       \\ \hline
\multicolumn{2}{c}{\cellcolor[HTML]{9B9B9B}\textbf{\begin{tabular}[c]{@{}c@{}}Energy Efficient Clustering \\ and Routing\end{tabular}}}                                         & {\color[HTML]{00009B} Addressed} & {\color[HTML]{00009B} Addressed} & {\color[HTML]{00009B} Addressed} & {\color[HTML]{00009B} Addressed}                                                       \\ \hline
\multicolumn{2}{c}{\cellcolor[HTML]{9B9B9B}\textbf{\begin{tabular}[c]{@{}c@{}}Sensor \\ Localization\end{tabular}}}                                                             & {\color[HTML]{00009B} Addressed} & {\color[HTML]{00009B} Addressed} & {\color[HTML]{00009B} Addressed} & {\color[HTML]{FE0000} Not Addressed}                                                       \\ \hline
\end{tabular}

\end{table*}

PSO, GA and ACO well address all the four problem domains of WSNs. Also, some hybrid techniques emerge for the same. Every new attempted claimed to show improved results over the previous approaches. In continuation of that, we have introduced the LO to solve the issues in WSNs. Table \ref{tab:present status}, shows the current status of all the four prominent algorithms. 

\section{Optimal Coverage using IGA-BACA and LO}
\label{Optimal coverage using IGA-BACA and LO}
  Getting optimal coverage in WSN belongs to a multi-objective optimization problem. The existing sensors, $N$, is represented by set ${S=(s_1,s_2,...,s_i,...,s_N)}$. In this optimization problem, we aim to estimate a sensor set ${S'}$, which covers the monitoring area to the maximum with minimum working sensors. The function for maximum coverage and minimum sensors is ${f_1(S')}$ and ${f_2(S')}$. Both these functions are conflicting in nature; undermining them both, the new objective function by changing it to a maximal objective function ${f(S')}$ read as;
  
    \begin{equation}\label{eq:9}
    \begin{aligned}
    \noindent
    \scalebox{0.9}{
    $ {f(S')=(f_1(S').f_1(S')/f_2(S'))}  $ }
    \end{aligned}
    \makebox[1.75cm]{}
    \end{equation}
    The framework for obtaining the optimal coverage using IGA-BACA and LO is explained in the upcoming subsections.
    
    \subsection{Mapping from Solution to Coding Space}
\begin{figure}
        \centering
        \includegraphics[width=0.5\textwidth]{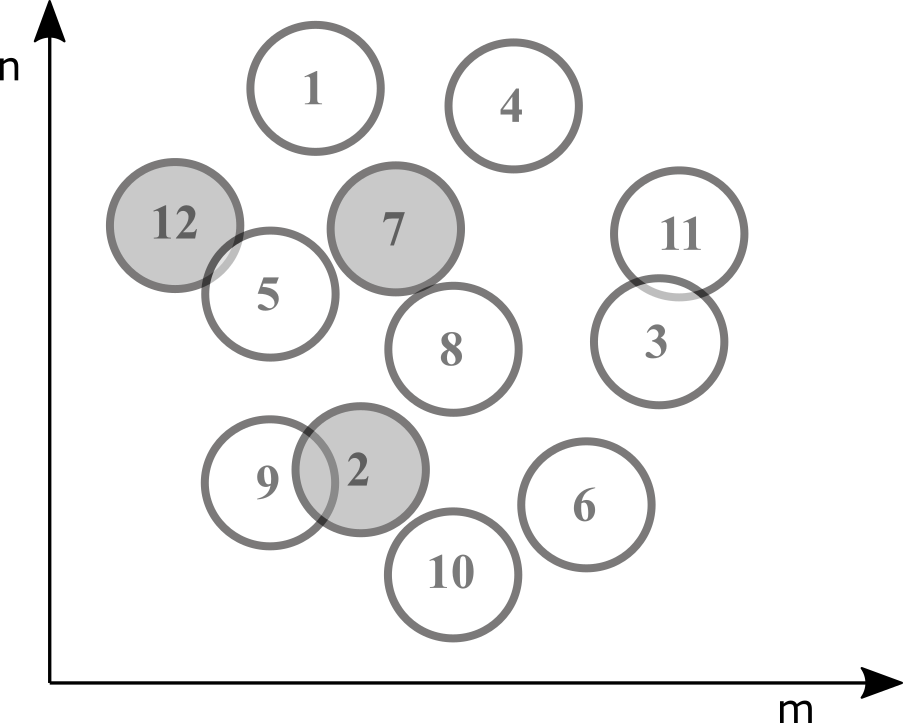}
        \caption{Illustrating a control vector with 9 active sensors out of 12 is  \big\{1 0 1 1 1 1 0 1 1 1 1 0\big\}. The dark circles represent the inactive sensors in the monitoring area. It is represented by a binary `0' in the control vector.}
        \label{f:2}
        \end{figure}

    The binary coding represents the position of the sensors in WSNs. The corresponding control vector is ${L=(l_1,l_2,...,l_i,...,l_N)}$. ${l_i}$ can either have a value of zero or one which represents the inactive or active state of a sensor respectively. The initialization of nomad and pride in LO and gene of the chromosome in GA has one to one correlation with the selection of nodes. Fig. \ref{f:2}, shows a typical example of a control vector. The probability of the sensor to be an active sensor depends on the adaptation or objective function (Equation \ref{eq:9}). Higher the value, the larger will be the probability.
    
\subsection{Algorithm and Process}

For optimization Equation \ref{eq:9}, we have used IGA-BACA and LO. In IGA-BACA, we have four processes, namely reproduction, crossover, mutation and update pheromone process. In contrast, LO has three processes, namely, mating, sorting, and elimination.

\begin{figure*}[ht!]
\centering
\includegraphics[scale=.7]{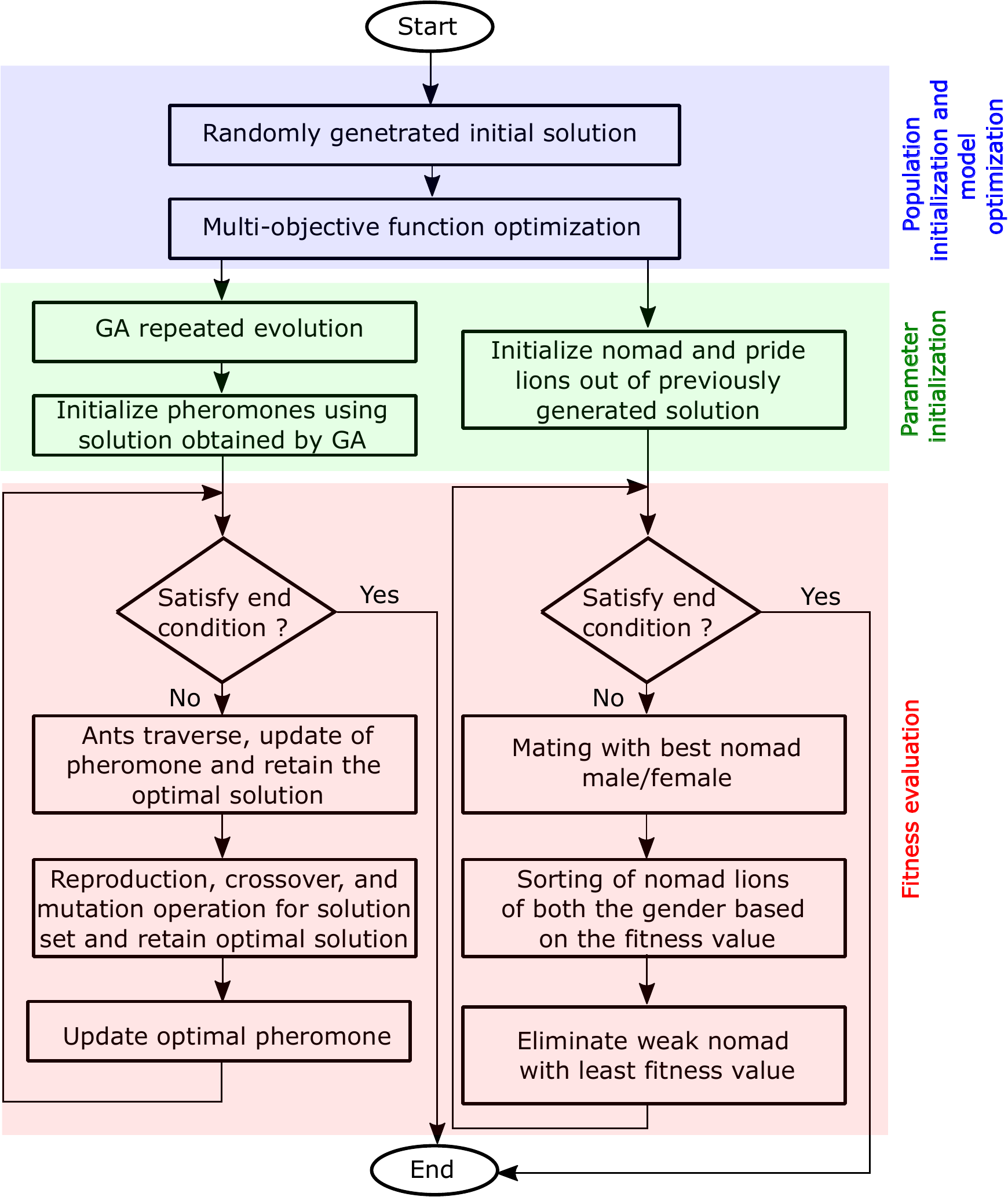}
\caption{Flowchart for IGA-BACA and LO.}
\label{f:Flow chart for IGA-BACA and LO}
\end{figure*}

In IGA-BACA, the first process reproduces new offspring depending on probability infraction to their fitness value. Afterwards, the new offspring are sorted based on their fitness values. The only offspring with high fitness values are retained, and others are discarded. This process ensures an increase in the average fitness of the colony. The only limitation associated with this process is that the number of possible varieties is lost. This limitation is subsequently overcome by the crossover and mutation process. In the crossover process, a pair of offspring is selected based on the probability, ${P_c}$. This step further increases the probability that crossed solutions may produce offspring with high fitness value. Afterwards, the mutation process alters an offspring based on the probability, ${P_m}$. This step explores the unexplored genetic material. Lastly, the update pheromone process (${(T_g)}$; pheromones update operator) mapped the updated pheromone for an optimal offspring elected by the ant sequence. The ant release the pheromones in the optimal path traversed by them using Max-Min rule. We can calculate the probability of the update pheromone operator by

     \begin{equation}\label{eq:10}
     \begin{aligned}
     \noindent
     \scalebox{1}{
     $  P \big\{T_g=x_i\big\}= \frac{f(x_i)}{ \sum^N_{k=1} f(x_k) }   $ }
     \end{aligned}
     \makebox[1.75cm]{}
     \end{equation}
 Where, ${N}$ is the number of offsprings.
 In LO, first mating with the best nomad, both male and female, is done followed by sorting nomad lions of both gender-based on fitness value. After which the nomad with least fitness value is eliminated. Analogous terms between the LO parameters and WSNs are listed in Table \ref{table}. The complete methodology for LO and IGA-BACA is illustrated in Fig. \ref{f:Flow chart for IGA-BACA and LO}.

\section{System Model}
\label{System Model}
As stated earlier, getting optimal coverage is one of the crucial problems associated with WSNs. The network should have maximum coverage with a certain level of QoS \citep{AMMARI20121935,Chen2013,BOUKERCHE201854}. Presence of blind area significantly affects the QoS threshold and network coverage rate, ultimately affecting the network reliability. In order to increase network reliability, we can deploy more sensors in critical areas. Increasing sensors will increase the network cost. In this paper, we used the bio-inspired algorithms to find the optimal node-set.

In this study, we have assumed that the total monitoring area, ${A}$, is a two-dimensional plane and It is split into $m$ $\times$ $n$ equal grids. After that, we have randomly distributed ${N}$ no. of sensors in the study area. Mathematically, these  sensors are represented by  ${S=(s_1,s_2,...,s_i,...,s_N)}$. All the sensors have effective radii (sensing radius) of ${r}$ with a coordinate ${(x_i,y_i)}$ for ${s_i}$.

In order to ensure maximum coverage, each grid in the monitoring area is considered as a target point. Mathematically, it is represented by $A={(a_1,a_2,...,a_j,...,a_{m x n})}$. If the target point  ${a_j}$ lies in the sensing region of sensor ${s_i}$, then the Euclidean distance between them is given by ${d(a_j,s_i) \leq r  }$ \citep{Tian2016}.

\begin{table*}[h]

\centering
\caption{Analogous mapping between LO algorithm and WSNs.}
\label{table}
\begin{tabular}{cc}
\hline
LO algorithm          & {Optimal coverage problem}                                                                                      \\ \hline
Solution of a food source       & Node distribution                                                                                \\ 
${N}$ dimensions in each solution               & ${N}$ sensor coordinates                                                                              \\ 
Fitness of the solution               & Coverage rate in $A$                                                                               \\ 
Maximum fitness         & Optimum deployment                                                                              \\ \hline
\end{tabular}
\end{table*}
The probability, ${P_{cov} (x,y,s_i )}$, that any coordinate ${(x,y)}$ in $A$ is sensed by a sensor ${s_i (x_i,y_i)}$ is given by
 \begin{equation}\label{eq:11}
             \begin{aligned}
             \noindent
             \scalebox{1}{
             $ P_{cov}(x,y,s_i)= \begin{cases}1 & (x-x_i)^2-(y-y_i)^2 \leq r^2 \\0 & otherwise\end{cases}   $ }
             \end{aligned}
             \makebox[1.75cm]{}
             \end{equation}
The area covered by the sensors is given by
 \begin{equation}\label{eq:12}
 \begin{aligned}
 \noindent
 \scalebox{1}{
 $ A_{area}(S)= \sum_{x=1}^m  \sum_{y=1}^n P_{cov}(x,y,s_i)\Delta x\Delta y        $ }
 \end{aligned}
 \makebox[1.75cm]{}
 \end{equation}
If $S'$ is the set of working or active sensors, then the fitness or objective function for the network coverage is given by 
 \begin{equation}\label{eq:13}
 \begin{aligned}
 \noindent
 \scalebox{1}{
 $ f_1(S')=A_{area}(S')/A_s$ }
 \end{aligned}
 \makebox[1.75cm]{}
 \end{equation}
 In contrast, the objective function for the node uses rate is given by 
 \begin{equation}\label{eq:14}
  \begin{aligned}
  \noindent
  \scalebox{1}{
   $ f_2(S')= |S' |/N        $ }
                         \end{aligned}
                         \makebox[1.75cm]{}
                         \end{equation}
Where ${N}$ is the total number of sensor nodes. Equation \ref{eq:13} and \ref{eq:14} are combined to form a  multi-objective optimization coverage problem given by.
\begin{equation}\label{eq:15}
\begin{aligned}
\noindent
\scalebox{1}{
$max f(S')=max( f_1(S'),1-f_2(S'))       $ }
\end{aligned}
\makebox[1.75cm]{}
\end{equation}
We have to maximise the equation \ref{eq:15} to get maximum coverage with minimum sensor node. 

\section{Simulation Results}
\label{Simulation Results}
The simulation parameters that we have used in this study is given in Table \ref{tablesimulation}. We selected a monitoring area ($A$) of ${ 100 }$ m $\times$ ${100 }$ m in which sensor nodes having a perception radius of 10 m ($r$ = 10 m) are deployed. The constants $k_1$, $k_2$, $k_3$, and $k_4$ are set to 1, 0.5, 1 and 0.5 respectively. These values restrict the range of $P_c$ between 0.5 and 1 (\textit{i.e.}, 0.5 $<$ $P_c$ $<$ 1) and $P_m$ between 0.001 and 0.05 (\textit{i.e.}, 0.001 $<$ $P_m$ $<$ 0.05). The moderately large range of $P_c$ and small range of $P_m$ is required for extensive recombination of solutions and prevention of the disruptions of the solutions respectively which ultimately prevents the algorithm from getting stuck into local optimum. The constant $\alpha$ controls the pheromone importance, while $\beta$ controls the distance priority. In general, $\beta$ should be greater than $\alpha$ for the best results. Both these parameters are interlinked. We have to fix one and vary (iterate) the other to find the optimal set of value. In this study, we fixed the $\alpha$ to 1 and found $\beta$ to be 6.

\begin{table}[h!]
\centering
\caption{Simulation parameters.}
\label{tablesimulation}
\begin{tabular}{cc}
\hline
\textbf{Parameter}                            & \textbf{Value}                    \\ \hline
Monitoring area (${A}$)                           & 100 m $\times$ 100 m \\
Perception radius ($r$)                         & 10 m                              \\
$N$                                             & 100                               \\
$k_1$ = $k_3$                    &  1              \\
$k_2$ = $k_4$ & 0.5
\\
$\alpha (\alpha \geq 1)$ & 1  \\
$ \beta ( \beta \geq 1)$ & 6
\\ \hline
\end{tabular}
\end{table}

 \begin{figure}
 \centering
\includegraphics[scale=0.5]{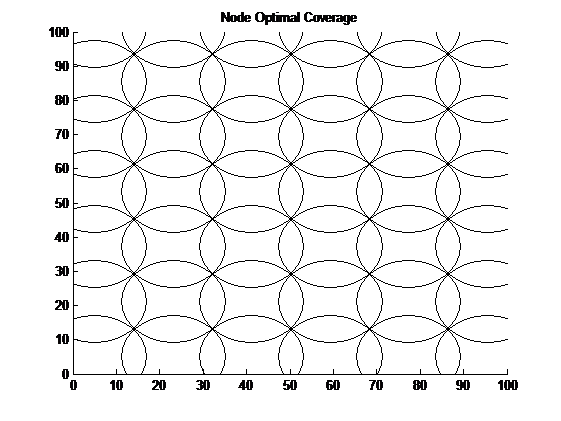}
\caption{Optimal network coverage.}
\label{f:4}
\end{figure}
    
 \begin{figure}
 \centering
\includegraphics[scale=0.5]{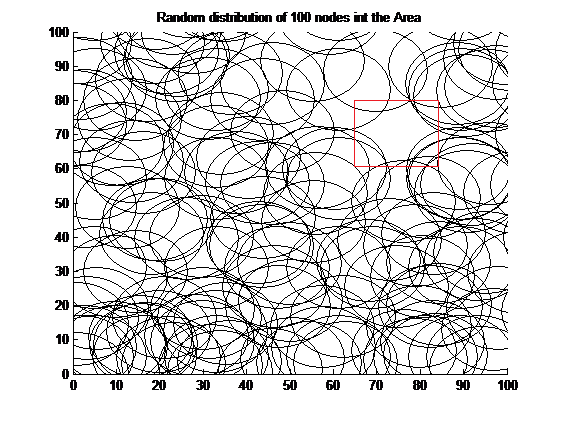}
\caption{Random distribution of 100 nodes.}
\label{f:5}
\end{figure}
We implemented the corresponding algorithm in MATLAB{\textsuperscript \textregistered} (version 2017b). We iterated the IGA-BACA algorithm for ${300}$ iterations. In doing so, we found that only ${42}$ (out of $100$) sensors cover the monitoring area optimally, as shown in Fig. \ref{f:4}. This optimal coverage is treated as a benchmark for further analysis. In contrast, while distributing these $100$ sensors randomly, we found a network coverage map, as shown in Fig. \ref{f:5}. We randomly distribute these ${100}$ sensors in the monitoring area, as shown in Fig. \ref{f:5}. Although the monitoring area is almost covered completely, there exist a significant amount of redundant nodes which senses redundant information. The uncovered area in the target monitoring area is considered as a coverage hole, and in Fig. \ref{f:5}, we can easily detect such coverage hole or blind areas (highlighted in red boxes). Hence, random network coverage is not usually adopted.
 \begin{figure}
 \centering
\includegraphics[scale=0.5]{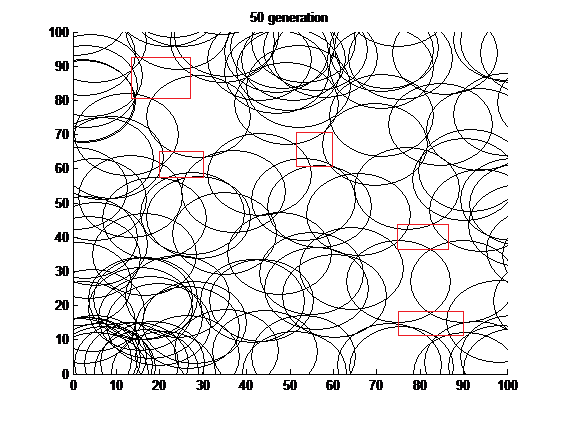}
\caption{50 Generation of IGA-BACA.}
\label{f:6}
\end{figure}
 \begin{figure}
 \centering
\includegraphics[scale=0.5]{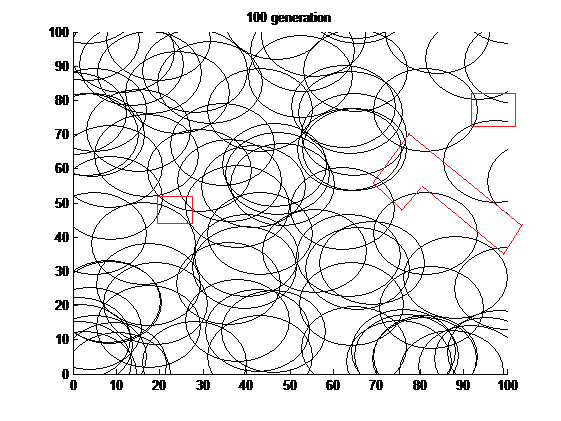}
\caption{100 Generation of IGA-BACA.}
\label{f:7}
\end{figure}
 \begin{figure}
 \centering
\includegraphics[scale=0.5]{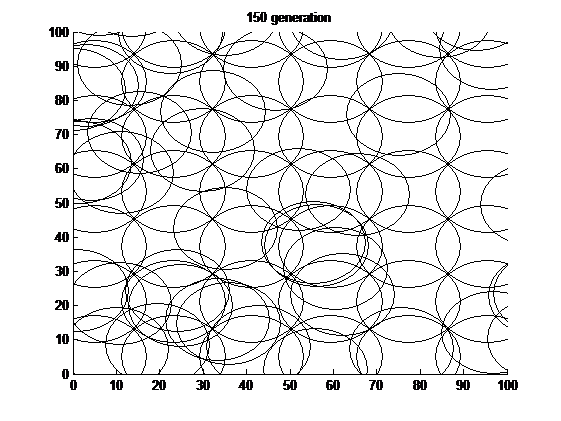}
\caption{150 Generation of IGA-BACA.}
\label{f:8}
\end{figure}

 \begin{figure}
 \centering
\includegraphics[scale=0.5]{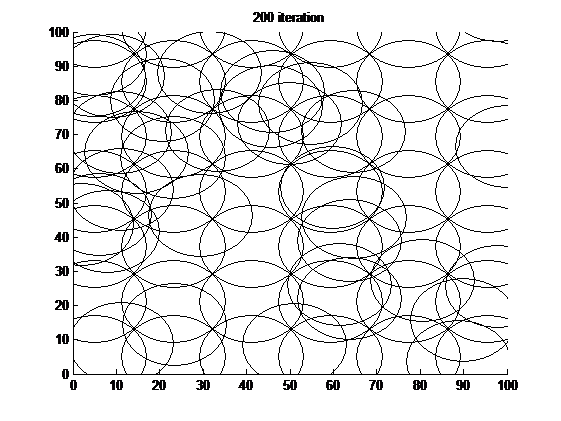}
\caption{200 Generation of IGA-BACA.}
\label{f:9}
\end{figure}
 \begin{figure}
 \centering
\includegraphics[scale=0.5]{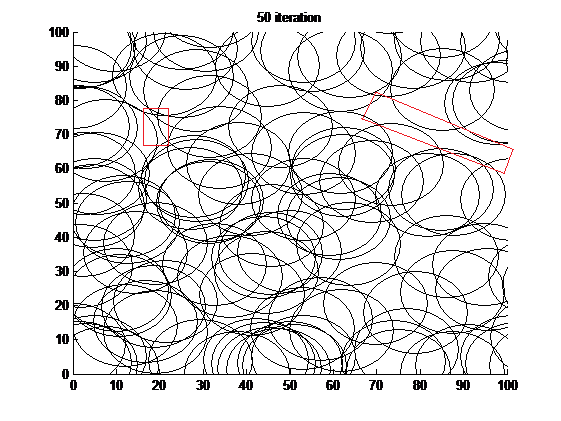}
\caption{50 Generation of LO.}
\label{f:10}
\end{figure}
 \begin{figure}
 \centering
\includegraphics[scale=0.5]{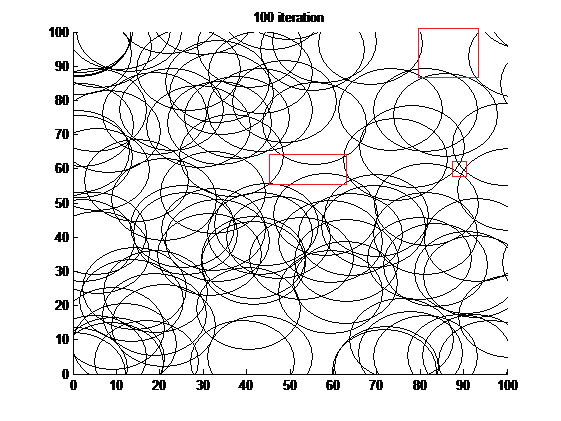}
\caption{100 Generation of LO.}
\label{f:11}
\end{figure}
 \begin{figure}
 \centering
\includegraphics[scale=0.5]{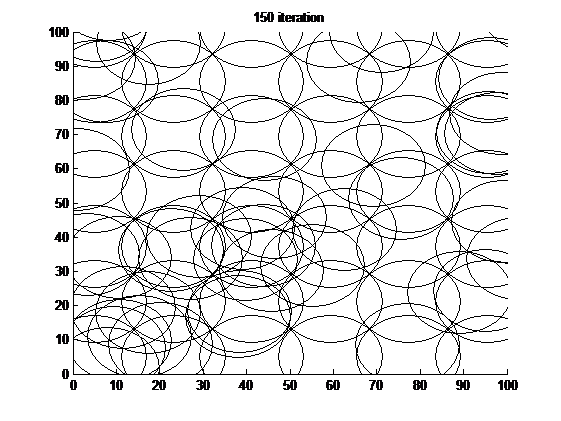}
\caption{150 Generation of LO.}
\label{f:12}
\end{figure}
 \begin{figure}
 \centering
\includegraphics[scale=0.5]{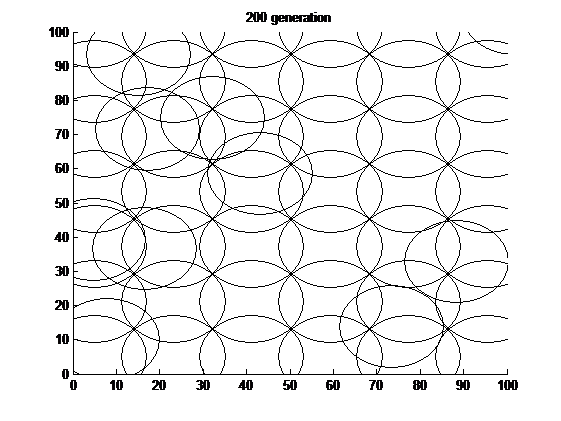}
\caption{200 Generation of LO.}
\label{f:13}
\end{figure}
 \begin{figure}
 \centering
\includegraphics[scale=0.5]{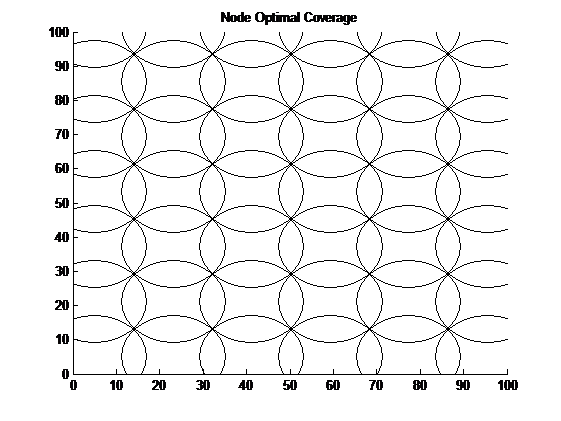}
\caption{250 Generation of LO.}
\label{f:14}
\end{figure}

 \begin{figure}
 \centering
\includegraphics[scale=0.5]{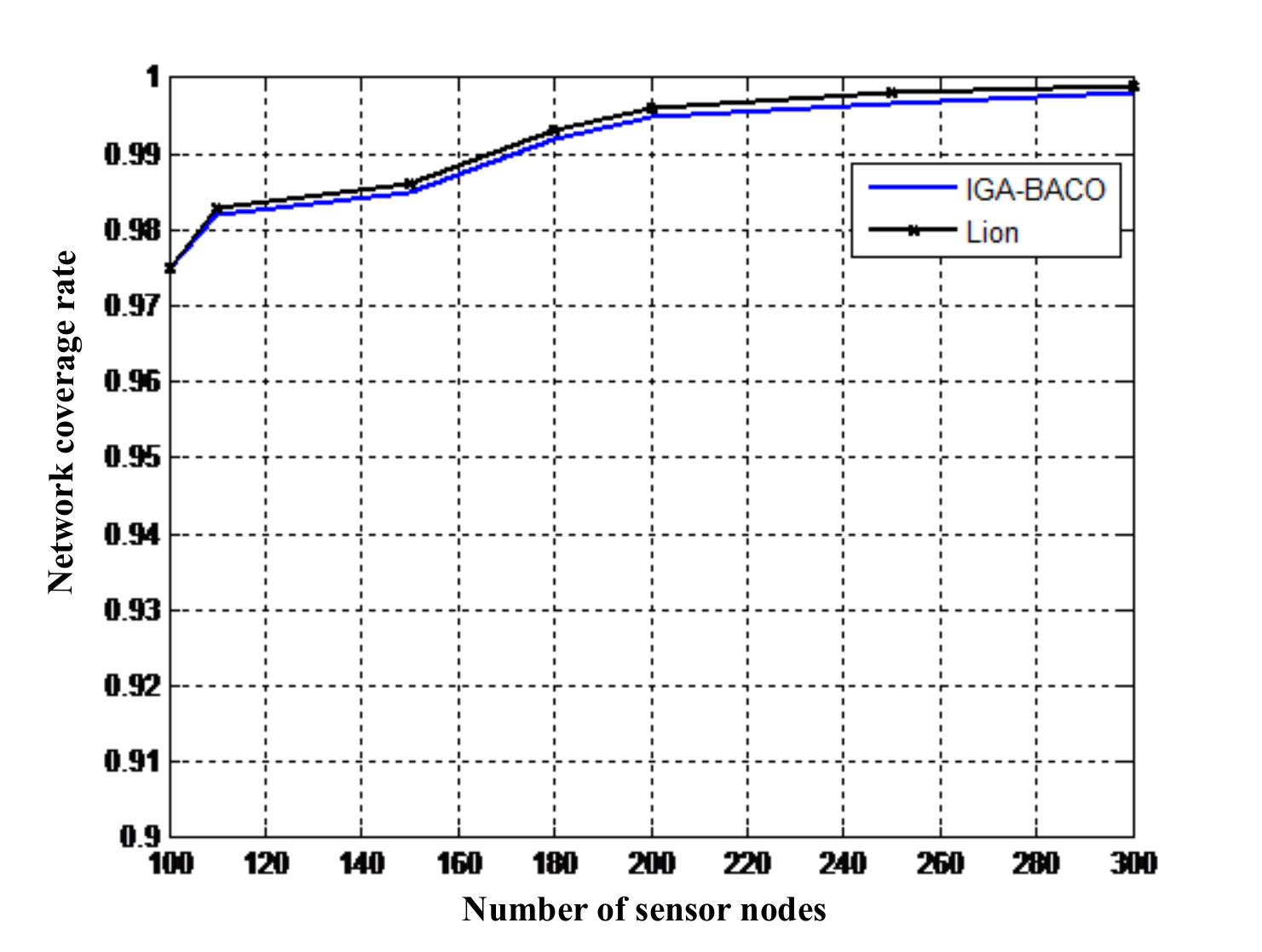}
\caption{Network coverage vs sensor.}
\label{f:15}
\end{figure}
 
 \begin{figure}
\centering
\includegraphics[scale=0.5]{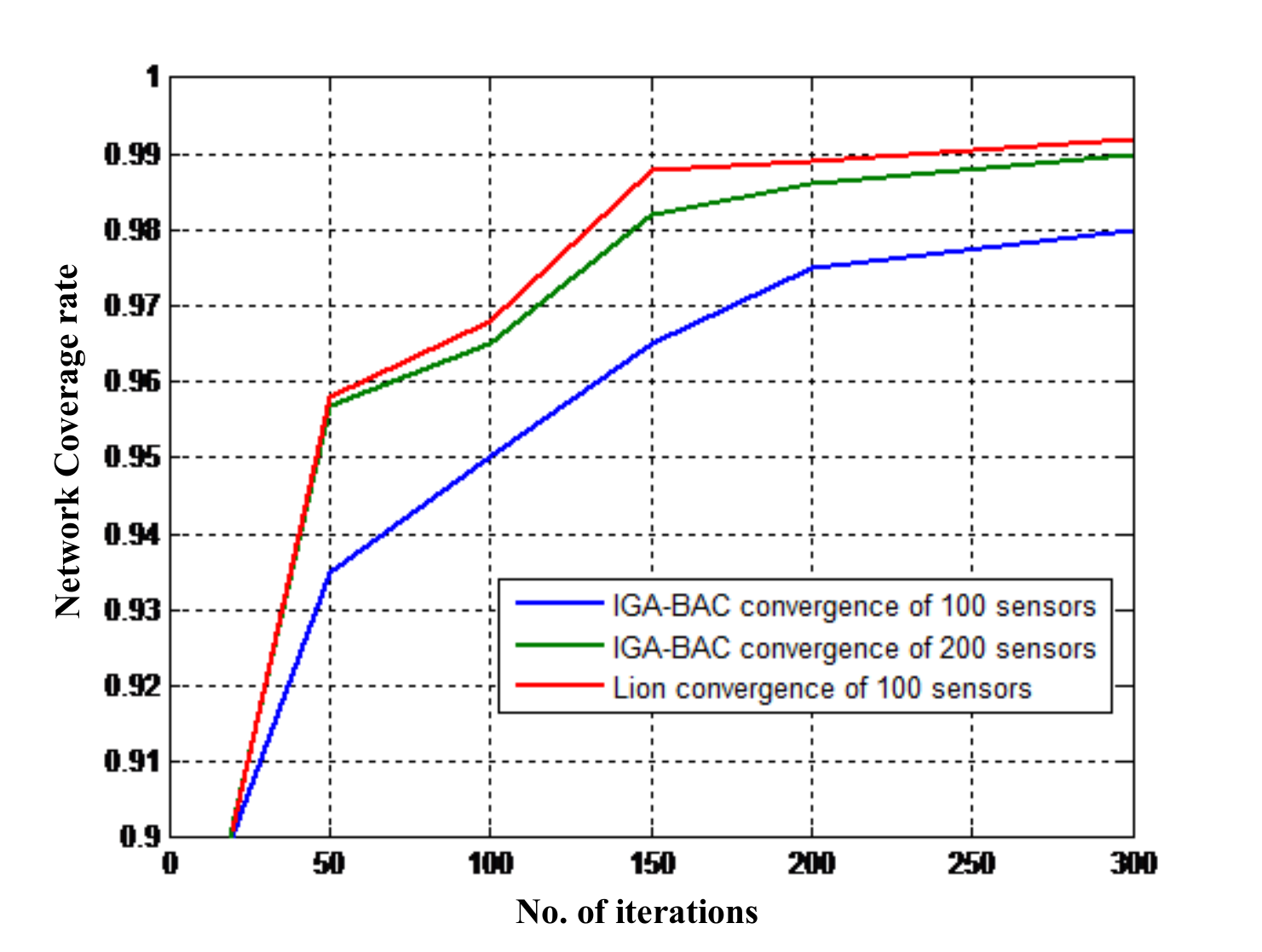}
\caption{Network coverage vs generation.}
\label{f:16}
\end{figure} 
\begin{table*}[h!]
\centering
\caption{Simulation results for IGA-BACA and LO.}
\label{tablesimulation_results}
\resizebox{1\columnwidth}{!}{
\begin{tabular}{ccccccc}
\hline
{\textbf{\begin{tabular}[c]{@{}c@{}} \\ Iterations\end{tabular}}\textbf{}} & \multicolumn{3}{c}{\textbf{IGA-BACA}} & \multicolumn{3}{c}{\textbf{LO}} \\ \cline{2-7} 
 & \textbf{\begin{tabular}[c]{@{}c@{}}Network coverage\\ rate (\%)\end{tabular}} & \textbf{\begin{tabular}[c]{@{}c@{}}Active sensor\\ node\end{tabular}} & \textbf{\begin{tabular}[c]{@{}c@{}}Time (s)\\ (GPU)\end{tabular}} & \textbf{\begin{tabular}[c]{@{}c@{}}Network coverage\\ rate (\%)\end{tabular}} & \textbf{\begin{tabular}[c]{@{}c@{}}Active sensor\\ node\end{tabular}} & \textbf{\begin{tabular}[c]{@{}c@{}}Time (s)\\ (GPU)\end{tabular}} \\ \hline
50 & 93.5 & 72 & 5 & 95.9 & 67 & 4 \\
100 & 95.0 & 65 & 11 & 96.9 & 61 & 9 \\
150 & 96.4 & 59 & 17 & 98.7 & 55 & 15 \\
200 & 97.5 & 53 & 22 & 98.9 & 48 & 19 \\
250 & 97.9 & 48 & 26 & 99.1 & 42 & 23\\ \hline
\end{tabular}
}
\end{table*}

Network coverage using the IGA-BACA algorithm for 50, 100, 150, 200 iterations (or generations) is shown in Fig. \ref{f:6} - \ref{f:9}. As we increase the number of iterations from 50 to 200,  we found that the network coverage tends towards the optimal coverage; hence the number of redundant nodes decreases significantly. In comparison with the IGA-BACA derived results, the network coverage using LO algorithm for 50, 100, 150, 200 and 250 iterations is shown in Fig. \ref{f:10} - \ref{f:14}. We found a similar trend of moving towards the optimal coverage with an increase in the number of iterations. For both the algorithms (\textit{i.e.}, IGA-BACA and LO), we have tabulated the network coverage rate (in percentage), GPU processing time (in seconds) and the number of active sensors corresponding to 50, 100, 150, 200 and 250 iterations as shown in Table \ref{tablesimulation_results}. We compared the results that are obtained through combined meta-heuristic IGA-BACA with the results obtained through LO. In doing so, we observed that the optimal network coverage is obtained by both the approach. However, IGA-BACA algorithm requires approximately $300$ iterations and LO algorithms require $250$ iterations to achieve the optimal coverage. Also, in LO, the optimal coverage is obtained at a lesser number of sensors, as shown in Fig. \ref{f:15}. Further, LO has a faster rate of convergence which is primarily due to the presence of a large number of local maxima with higher values of fitness functions (Table \ref{tablesimulation_results}). In addition to this, we plotted the network coverage rate against the number of iteration (Fig. \ref{f:16}). In doing so, we observed that the network coverage increases as the function of iteration. Also, the convergence of LO is better than IGA-BACA.
    
\section{Conclusion}
\label{Conclusion}
The use of nature-inspired algorithms has created a new era in next-generation computing. These algorithms are well suited for solving multi-objective optimization problems. Various features of nature-inspired algorithms such as reasonable computational time, find global optimal and applicability make them well suited for real-world optimization problems. In contrast, traditional algorithms generally fail to provide satisfactory results mainly because of the complexity and size of the problem structure.  In this paper, we have presented a comprehensive review of such algorithms in context to various issues related to the WSNs.

We have evaluated the potential of two efficient meta-heuristic approaches that compute the optimal coverage in WSNs, namely IGA-BACA and LO. We have compared the results of both these approaches. In doing so, we observed that as the number of iteration is increasing the network coverage rate tend towards optimal coverage. Also, the network coverage rate is faster in LO approach as compared to IGA-BACA. The optimal coverage is achieved with a lesser number of iteration in case of LO as compared with other approaches. It is due to the presence of a large number of local maxima with higher fitness value, and hence it is hardly any chance to miss local maxima. Although LO gives better performance than other optimization algorithms, still there is much scope to explore this algorithm and to apply it in multi-objective problems. For instance, if we can use machine learning approach such as Artificial Neural Network (ANN) that incorporates combined heuristic such as Ant Lion Optimization (ALO), IGA-BACA, etc. as our system inputs.

\section*{Conflict of Interest}
The author states that there is no conflict of interest. 

\section*{CRediT author statement}
Abhilash Singh and Sandeep Sharma: Conceptualization, Methodology, Software.
Abhilash Singh and Sandeep Sharma: Data curation, Writing- Original draft preparation, Visualization, Investigation. Sandeep Sharma: Supervision. 
Abhilash Singh, Sandeep Sharma and Jitendra Singh: Software, Validation.
Abhilash Singh, Sandeep Sharma and Jitendra Singh: Writing- Reviewing and Editing.

\section*{Acknowledgment}

We would like to acknowledge IISER Bhopal, Gautam Buddha University Greater Noida, and IIT Kanpur for providing institutional support. We thank to the editor and all the anonymous reviewers for providing helpful comments and suggestions.
\bibliography{mybibfile}

\end{document}